\pdfoutput=1	% To force pdflatex processing on arXiv
\documentclass[9pt,column,twoside,conference]{IEEEtran}
\usepackage{framed}
\usepackage{moreverb}
\usepackage[small,sf,bf]{subfigure}
\usepackage{xspace}
\usepackage{flushend}			%

\usepackage{xfrac}			%

\usepackage{amsfonts}
\usepackage{amssymb}
\usepackage{amsmath}
\usepackage[thmmarks,amsmath,hyperref]{ntheorem}
\usepackage{graphicx}
\usepackage{times}
\usepackage[scaled]{helvet} 		%
\usepackage{textcomp}
\usepackage{color}
\usepackage{colortbl}
\usepackage{booktabs}
\usepackage{listings}
\usepackage{multicol}
\usepackage{times}
\usepackage{pifont}			%
\usepackage{microtype}

\usepackage{mmm}
\usepackage{wand-sample}
\usepackage[scaled=0.85]{beramono}
\usepackage[T1]{fontenc}
\usepackage{url}
\usepackage{multirow}
\usepackage{array}

\definecolor{listinggreen}{rgb}{0,0.6,0}
\definecolor{listinggray}{rgb}{0.5,0.5,0.5}
\definecolor{listingmauve}{rgb}{0.58,0,0.82}
\definecolor{listingkeywordcolor}{rgb}{1.0,0.4,0.0}
\definecolor{listinglightgray}{rgb}{0.9863,0.9863,0.9863}

\lstdefinelanguage{FSharp}
{
	morekeywords	= {
    let,
    type,
    Measure,
	},
	sensitive	= false,
	morecomment	= [l]{\#},
	morecomment	= [s]{(*}{*)},
}

\lstdefinelanguage{Newton}
{
	morekeywords	= {
		signal,
		derivation,
		symbol,
		name,
		invariant,
		constant,
		English,
		sensor,
		name,
		none,
		dot,
		cross,
		derivative,
		integral,
		interface,
		i2c,
		spi,
		analog,
		write,
		read,
		delay,
		range,
		erasuretoken,
		uncertainty,
		accuracy,
		precision,
		Gaussian,
		exponential,
		biexponential,
		to,
		bits,
		dimensionless,
		include
	},
	sensitive	= false,
	morecomment	= [l]{\#},
	morecomment	= [s]{/*}{*/},
}

\usepackage{hyperref}
\hypersetup{
    colorlinks=false, %
    linktoc=all,     %
    linkcolor=blue,  %
}

\usepackage[bordercolor=shadeOne, backgroundcolor=shadeTwo, linecolor=red, textwidth=0.8in, textsize=tiny]{todonotes}
\usepackage{ulem}
\usepackage{float}

\definecolor{shadeOne}{rgb}{0.9,0.95,0.95}
\definecolor{shadeTwo}{rgb}{0.99,0.99,0.99}

\begin{document}

\title{Bridging the Band Gap: What Device Physicists Need to Know About Machine Learning}

% Use letters for affiliations, numbers to show equal authorship (if applicable) and to indicate the corresponding author
\author{\IEEEauthorblockN{Nathaniel J. Tye\\}
	\IEEEauthorblockA{\textit{Cambridge Graphene Centre,} \\
		\textit{Department of Engineering,} \\ 
		\textit{University of Cambridge}\\
		njt48@cam.ac.uk \\}
	\and
	\IEEEauthorblockN{Stephan Hofmann \\}
	\IEEEauthorblockA{\textit{Department of Engineering} \\
		\textit{University of Cambridge}\\
		sh315@eng.cam.ac.uk \\}
	\and
	\IEEEauthorblockN{Phillip Stanley-Marbell \\}
	\IEEEauthorblockA{\textit{Department of Engineering} \\
		\textit{University of Cambridge}\\
		phillip.stanley-marbell@eng.cam.ac.uk}
}

\vspace{-0.1in}
%\keywords{Programming Languages $|$ Sensors $|$ Computer Architecture.}

%\dates{This manuscript was compiled on \today}
%	\doi{\url{www.pnas.org/cgi/doi/10.1073/pnas.XXXXXXXXXX}}

% Optional adjustment to line up main text (after abstract) of first page with line numbers, when using both lineno and twocolumn options.
% You should only change this length when you've finalised the article contents.
%\verticaladjustment{-5pt}
%\twocolumn[
%\begin{@twocolumnfalse}
\maketitle
\begin{abstract}
This article surveys the landscape of semiconductor materials and devices research for the acceleration of machine learning (ML) algorithms. We observe a disconnect between the semiconductor and device physics and engineering
communities, and the digital logic and computer hardware architecture communities. The article first provides an overview of the
principles of computational complexity and fundamental
physical limits to computing and their relation to physical
systems.

The article 
then provides an introduction to ML by presenting three key components of ML systems: 
\textit{representation}, \textit{evaluation}, and \textit{optimisation}. The article 
then discusses and provides examples of the application of emerging technologies from the semiconductor and device physics domains as solutions to
computational problems, alongside a brief overview of emerging materials and devices for computing applications.

The article then reviews the landscape of
ML accelerators, comparing fixed-function and reprogrammable digital logic with novel devices such as memristors, resistive memories, magnetic memories, and
probabilistic bits. We observe broadly lower performance of ML accelerators based on novel devices and materials when
compared to those based on digital complimentary metal-oxide semiconductor (CMOS) technology, particularly in the MNIST optical character recognition task, a common ML benchmark, 
and also highlight the lack of a trend of progress in approaches based on novel materials and devices.

Lastly, the article proposes figures of merit for meaningful evaluation and 
comparison of different ML implementations in the hope of fostering a dialogue 
between the materials science, device physics, digital logic, and computer architecture communities 
by providing a common frame of reference for their work.
\end{abstract}

%\end{@twocolumnfalse}]
%\thispagestyle{firststyle}
%\ifthenelse{\boolean{shortarticle}}{\ifthenelse{\boolean{singlecolumn}}{\abscontentformatted}{\abscontent}}{}

% If your first paragraph (i.e. with the \dropcap) contains a list environment (quote, quotation, theorem, definition, enumerate, itemize...), the line after the list may have some extra indentation. If this is the case, add \parshape=0 to the end of the list environment.
% \section{Introduction}

\begin{figure*}[h]
	\centering
	\includegraphics[width=\textwidth]{"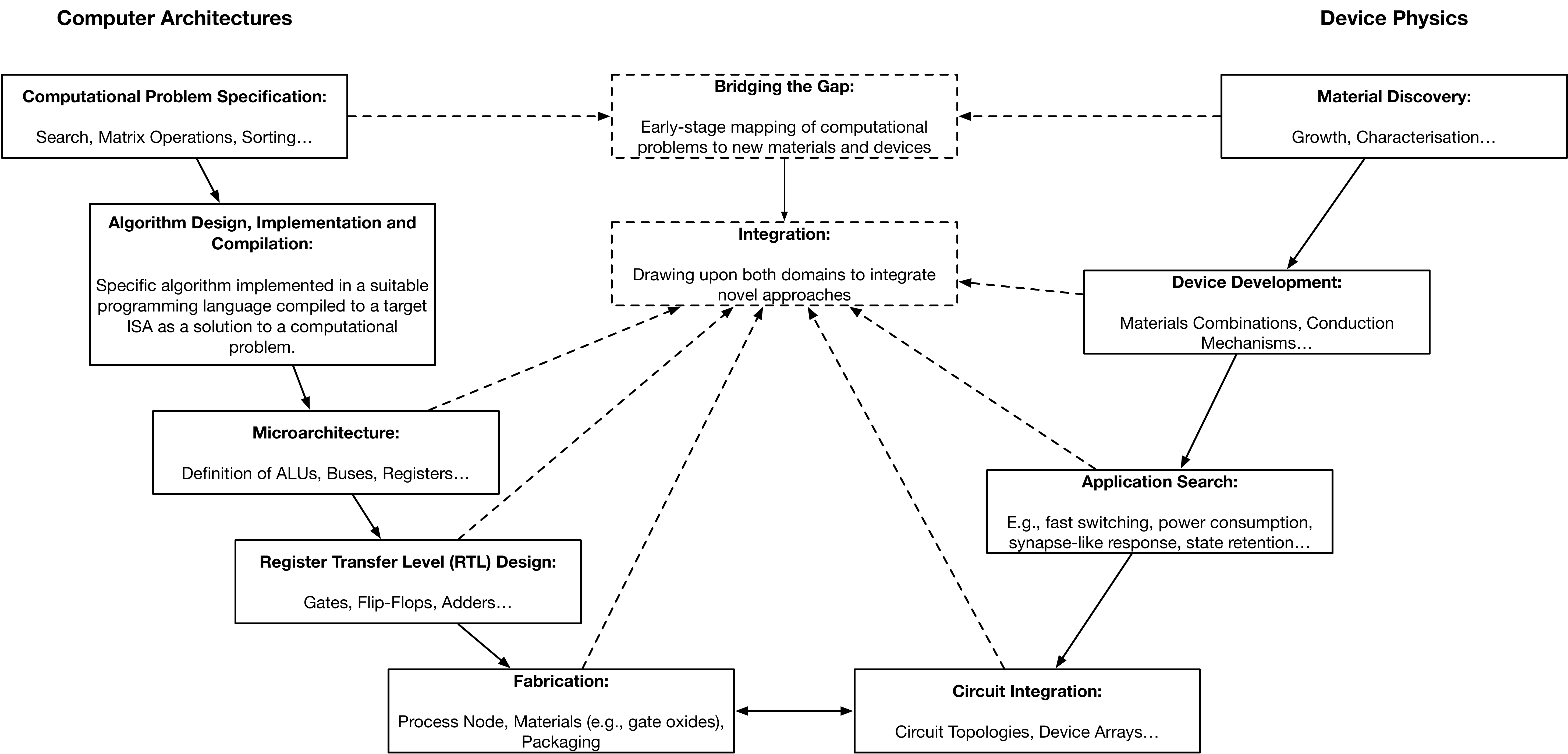"}
	\caption{Illustration of the development process for a computer architect compared to that of a device physicist. At present, the two only intersect at the bottom of each (Fabrication and Circuit Integration). We show the typical process with solid lines, and our proposal for new strategies with dotted lines. By intersecting earlier, e.g., at the first step, by mapping computational problems to devices and materials, each step in each community can be brought together, producing more impactful research.}
	\label{figure:DevProcesses}
\end{figure*}

\section{Introduction}
\label{section:introduction}

From the employment of graphical methods by ancient Babylonian astronomers to calculate Jupiter's position in the sky~\cite{Ossendrijver2016}, to the \textit{Antikythera mechanism}, used by ancient Greek navigators to track the paths of celestial bodies~\cite{Freeth2014}, humans have been doing computations for millennia. Computers are now an integral part of daily life, with applications from healthcare to construction, transport to fitness, agriculture to science. In the tradition of naming historical periods after dominant materials in tools, some claim that we are living in the `silicon age'~\cite{Chabal2001}~\cite{Siffert2013}. Silicon is the dominant material used in contemporary computing devices. The success of silicon and in particular, metal-oxide-semiconductor field-effect-transistors (MOSFETs), has been described and explained many times and in great detail, but to briefly summarise:

\begin{enumerate}
	\item Stable Oxide Formation: Silicon forms a stable oxide layer which serves as a good semiconductor-insulator interface.
	\item Abundance: Silicon is the second most abundant element in the earth's crust, after oxygen, making up 28.2\% of its mass~\cite{Rumble2019}.
	\item Miniaturisation and Mass Production: MOSFETs are the first transistor which could be miniaturised and mass-produced~\cite{Moskowitz2016}.
	\item Power Efficiency: Complimentary metal-oxide-semiconductor (CMOS) eliminates the always-on power draw of n- or pMOS by combining both into a single device which only consumes power when the device switches.
\end{enumerate}
However, physical constraints and new application classes are demonstrating the limitations of conventional computing methods. The growing prevalence of machine learning (ML) and big data, alongside the diminishing returns and increasing difficulty of CMOS scaling has triggered a search for alternative and complimentary approaches to computation. ML training compute demands have doubled roughly every 3.4 months since 2012~\cite{amodei_2021}, significantly faster than the roughly two year doubling time of integrated circuit transistor counts described by Moore's Law~\cite{Moore1965}. Despite a large research focus, very few of these new technologies have seen adoption in mainstream computing systems. This article examines and addresses the varied reasons for this, with a focus on ML hardware.

\subsection{Definitions}
We provide the following definitions of common terms which will be used in this article.

\begin{enumerate}
	\item Algorithm: a sequence of operations used to perform a computation, independent of implementation.
	\item Computer Architecture: an abstract description of a computer, such as an instruction set architecture (ISA), which defines functionality but not the physical implementation. See Section~\ref{section:architectures} for a more thorough explanation.
	\item Computer Device: an individual device used to represent data for computation, e.g., a transistor, or a resistive random-access memory (ReRAM).
	\item Software: An implementation of an algorithm or algorithms generally intended to be executed on reprogrammable digital CMOS hardware (e.g., a microprocessor).
	\item Neuromorphic System: a hardware device or circuit which mimics the behaviour of biological neural systems. See our discussion in Section~\ref{section:NeuroSys} for a more detailed exploration of the term.
	\item Fixed-Function Digital Hardware: a device or circuit which processes data using digital logic in a fixed manner. For example, an H.264 decoder is an application-specific integrated circuit (ASIC) designed to decode video streams encoded in a common format.
	\item Programmable Architectures: a piece of hardware that can be reprogrammed for arbitrary functionality, such as a microprocessor. Such a system reuses the same logic elements for different applications.
	\item Hardware Microarchitecture: the physical specification of an ISA, defining ALUs, buses, registers, etc.	
\end{enumerate}

\subsection{Bridging the Gap}
For continued development of computer technology, significant and meaningful advances in computation must be made on two main fronts: computational devices and computer architectures. However, the disconnect between the devices and computing communities may pose a greater hurdle to progress than technological challenges. Although it will be necessary to examine the state of the art in the research areas we discuss in this article, we do not intend this article to be a literature review of all the topics touched upon. A number of comprehensive review articles across the fields discussed here already exist for the interested reader~\cite{Wolpert2019}~\cite{Tanaka2019}~\cite{Sangwan2020}~\cite{Li2018}~\cite{Seok2014}. In the following sections, this article explores the main areas of active research and identifies and offers solutions to the disconnect we observe between the devices, materials and computing communities. Our hope is that this will facilitate a better understanding between the communities and stimulate dialogue, collaboration, and progress.

Successful examples of machine learning hardware development processes serve as a useful way to observe a typical development cycle, as well as where issues may arise. Norrie et al.~\cite{Norrie2021} provide such an example, describing the development of the v2 and v3 versions of Google's Tensor Processing Unit (TPU). Figure~\ref{figure:DevProcesses} illustrates the typical development process for a new integrated circuit from the perspectives of a computer architect and a device physicist. We see that, for the former, this begins by identifying a specific computational problem (see Jongerius et al., for an overview of contemporary computational problems~\cite{Jongerius2011}), an implementation of a specific algorithm, and then a survey of existing technologies to arrive at a given implementation, in essence, a top-down approach. For the latter, this begins with materials discovery, followed by device fabrication, and then integration of devices into a larger circuit. This results in a bottom-up approach, or, in practise, a solution looking for a problem. By examining the key considerations described by Norrie~\cite{Norrie2021}, we begin to see what considerations computer architects make when searching for a suitable solution. Norrie et al. divide tasks into two groups: those which need to be done well, and those which can be done to a working level. The first group, tasks which must be done well, consist of the following:

\begin{enumerate}
	\item Build Quickly: Fast design and fabrication of hardware, e.g., using readily-available components and `good-enough' designs;
	\item Achieve High Performance: High-bandwidth buses and memory, systolic array structure for high computation density, use of the \texttt{bfloat16} format, instruction-level parallelism;
	\item Scale Efficiently: Ability to add multiple processors to cope with increasingly-complex problems and datasets;
	\item Easily Adapt to New Workloads: \texttt{bfloat16} is easy for ML software to use, chip developed alongside compiler team to ensure programmability, the terminology of linear algebra used due to it's generality;
	\item Cost Effective: The systolic array structure allows high density without demanding a large chip area, \texttt{bfloat16} reduces hardware and energy costs, dual-core design, compiler-controlled memory hierarchy.
\end{enumerate}

Considering novel devices as solutions to computational problems is a key motivation of this article and provides a way to judge the most effective application of novel materials and devices to acceleration of ML computations, thus facilitating their consideration by computer architects. Table~\ref{table:CompProbsAndSols} gives some examples of existing uses of hardware as solutions to computational problems, grouping devices as potential solutions to computational problems. For example, artificial neural network (ANN) inference as a computational problem is simply matrix-vector multiplication (MVM), with ReRAM crossbars being a potential solution.

\begin{table}[h]
	\caption{Example mapping of computational problems to hardware solutions. Abbreviations: ReLU (rectified linear unit); GPU (graphics processing unit), DNN (deep neural network); SRAM (static random-access memory).}
	\centering
	\begin{tabular}{c|c|c}
		\toprule
		\bf{Computational}	 & \bf{Solution(s):}	& \bf{Applications:} \\
		\bf{Problem:} & \bf{} & \bf{} \\
		\midrule
		Matrix-Vector & ReRAM Crossbar~\cite{Tsai2018}, & DNN Inference, \\
		Multiplication & FPGA~\cite{Kestur2012},  &  \\
		& Google TPU~\cite{Norrie2021} & \\
		& & \\
		ReLU & ReRAM~\cite{Oh2021} & Activation Functions \\
		& & \\
		Differential & Neurotransistors (e.g.,  & SNNs, \\
		Equations & Integrate and Fire)~\cite{Wang2018},  &  DNN Training \\
		 & IBM TrueNorth~\cite{TrueNorth} & \\
		Binary Stochastic & P-Bits~\cite{Pagliarini2020} & Bayesian Networks \\
		Neuron & & \\
		& & \\
		Monte Carlo & SRAM~\cite{Kang2018} & Random Forests \\
		Cross-Validation & & \\
		\bottomrule
	\end{tabular}
	\label{table:CompProbsAndSols}
\end{table}

\section{Computational Complexity, Thermodynamics and Fundamental Limits}
\label{section:complexity}

\subsection{Computational Complexity}
Every computer algorithm, from the simple addition of two numbers to the Fourier transform, requires a certain amount of time and resources to run. This complexity is an important consideration in any computing system as it has implications for the efficiency and scalability of an algorithm's implementation. Complexity is generally evaluated in two main ways: time and space~\cite{Arora2009}~\cite{Goldreich2008}~\cite{Papadimitriou1994}.

\begin{itemize}
	\item Time complexity is the number of steps for some target computational model that an algorithm takes to run and is implementation independent. 
	\item Space complexity describes the memory cost of an algorithm. A given program requires temporary space during the execution of an algorithm (auxiliary space) and memory to store the input variable (input space). The total space complexity is the sum of these: Space Complexity = Auxiliary Space + Input Space.
\end{itemize}

\subsection{Big-O Notation}
Big-O notation is a way of describing algorithmic complexity as a function of input size according to the growth of time and space requirements. Formally, big-O notation describes the asymptotic upper bound of the function's complexity to within a constant factor~\cite{Cormen2009}.

Determining the complexity of a function is relatively simple: one simply selects the fastest growing term, ignoring constant terms. For example, if a function $f(x)$ is defined as $f(x) = an + b$, where $a$ and $b$ are constants and $n$ is the input, $a$ and $b$ can be ignored as they become small compared to $n$, when $n$ tends towards infinity. The big-O complexity of the function is thus $\mathcal{O}(n)$. The addition of numbers is an example of an algorithm with complexity $\mathcal{O}(n)$. Now consider another function, $g(x)$, defined as $g(x) = an^2 + bn + c$. This is a simple quadratic equation. By inspection, it is clear that, as $n$ tends to infinity, the $n^2$ term will grow larger more quickly than the $n$ term and so the big-O complexity of this function will be $\mathcal{O}(n^2)$. Traditional `long multiplication' has complexity $\mathcal{O}(n^2)$, however algorithms have been developed which reduce the time complexity of multiplication to $\mathcal{O}(n\log(n))$)~\cite{HHMultiplication}. 

\begin{figure}
	\centering
	\includegraphics[trim= 4cm 7.3cm 5cm 8cm, clip=true, width=\linewidth]{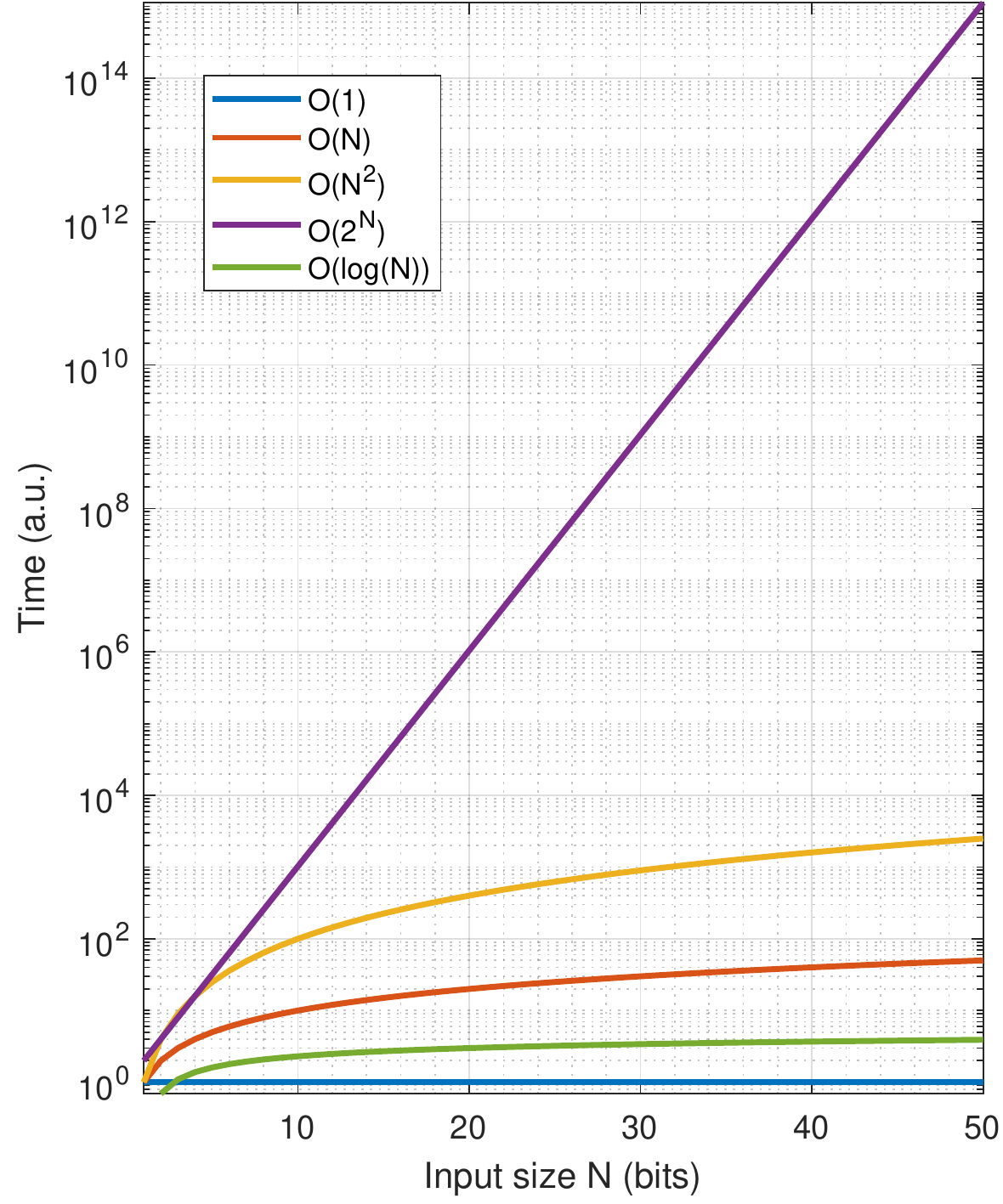}
	\vspace{-0.5cm}
	\caption{Computation time as a function of input size, $N$, for some standard complexities.}
	\vspace{-0.5cm}
	\label{figure:complexities}
\end{figure}

Though the actual time and resources required by an algorithm are dependent on the system running it, algorithmic complexity is always an important consideration. Consider two sorting algorithms: \textit{bubble sort}, with average-case complexity $\mathcal{O}(n^2)$ and \textit{merge sort}, with average-case complexity $\mathcal{O}(n\log(n))$. If one wishes to sort an array containing a million numbers, using both algorithms, with the bubble sort running on a computer that can complete a hundred million operations per second and the merge sort running on a slower computer which can only complete one million operations per second, then the time taken for the bubble sort is:

\begin{align*}
\frac{(10^6)^2}{(10^8)} = 10,000 s,
\end{align*}
whereas the slower computer, using merge sort, takes:

\begin{align*}
\frac{(10^6)\log(10^6)}{(10^6)} = 6 s.
\end{align*}
Even though the first computer is a hundred times faster, it performs the same operation significantly slower. This illustrates the importance of algorithmic complexity in a computing system. A fast algorithm on a slow implementation will perform better than a slow algorithm on a fast implementation. This observation is important regardless of whether the one or both of the implementations of the algorithm are achieved using a programmable processor, fixed-function digital hardware such as a digital ASIC or analog domain computation. Figure~\ref{figure:complexities} shows the time complexity growth of different big-O complexities with problem size.

\subsection{Precision in Computing}
In digital computing, the number of bits used to represent a number gives the precision of that number. Greater precision is generally a desirable goal. Modern ML implementations in fixed-function hardware typically make use of the \texttt{bfloat16} format~\cite{Norrie2021}, as it allows for fast conversion to 32 bit floating point formats whilst reducing storage requirements and computation time for ML algorithms. The reduced size allows for a sufficient level of accuracy to be retained whilst reducing the input size. For a 7\,nm process, Norrie et al. estimate a $1.5\times$ reduction in energy consumption for the \texttt{bfloat16} format compared to a conventional IEEE 16-bit float format.

However, more bits cost time, resources and energy. These constraints have led to a renewed interest in  computation, which, in theory, can offer infinite precision, as one does computation using continuous rather than discrete signals. In Section~\ref{section:computing-devices}, we discuss how analog approaches can offer significant savings in terms of resources and energy requirements. 

To readout or otherwise interact with the data one is computing involves measuring a signal. Thus we see that a truly analog computation is not possible, as measurement is a discrete process; an truly analog signal would require infinite precision. Analog systems are more sensitive to noise and offset than digital systems. As digital circuits rely on some discrete, reference state, the continuous nature of analog circuits means these cannot exist in analog systems~\cite{Sarpeshkar1998}. As computations become increasingly complex, the associated noise accumulation becomes a dominant and limiting factor in the precision. Compensating circuitry can of course be added to attenuate the noise, however this adds additional resources and energy costs to the system which may eventually eliminate any advantages in energy and space efficiency  afforded by an analog system.

\subsection{Physical Limits of Computation}
As well as mathematical constraints on computation, there also exist physical limits. Computation is ultimately a physical action and so obeys the laws of physics. One obvious limit is the speed of light and the resulting impossibility of faster-than-light communication. An exception that may be raised here is quantum entanglement, where two or more particles interact such that their quantum states are interdependent, and a measurement of one particle is perfectly correlated with a measurement of another. The entanglement remains regardless of the separation between the particles. However, this cannot be used for the purposes of communication or transmission of information as it would require a measurement result to be transmitted over a classical channel, as, without this, the measurement of one conveys no information~\cite{nielsen_chuang_2010}.

As information is a physical quantity~\cite{Shannon1962}, there are thus thermodynamic considerations. Wolpert provides a comprehensive overview of these~\cite{Wolpert2019}.

\subsubsection{The Classical Landauer Limit}
\begin{figure}[h]
	\centering
	\includegraphics[trim= 5cm 7cm 5.5cm 9cm, clip=true, width=0.8\linewidth]{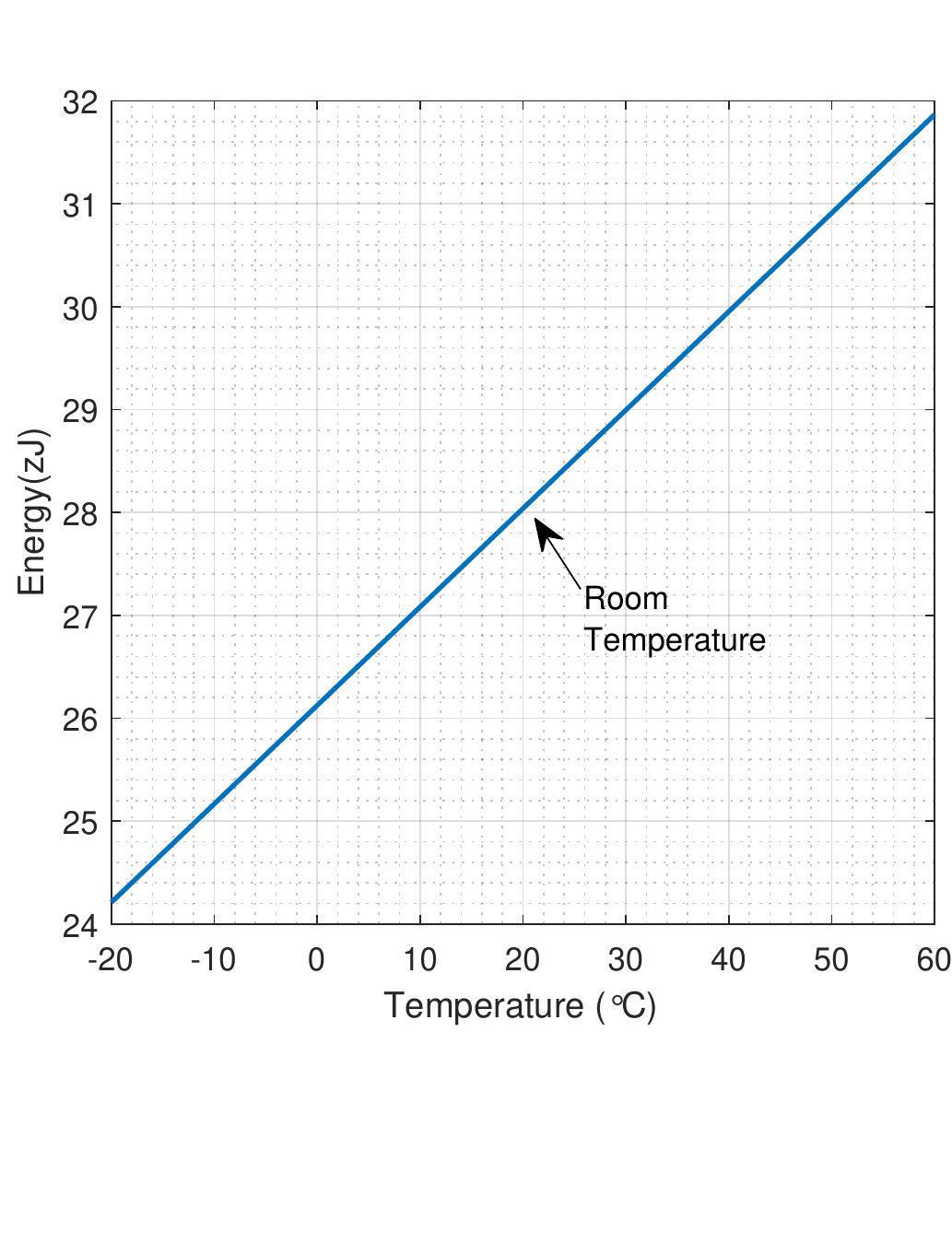}
	\vspace{-0.7in}
	\caption{A plot of the classical Landauer limit as a function of temperature.}
	\vspace{-0.5cm}
	\label{figure:Landauer}
\end{figure}

The Landauer limit asserts that there is a minimum energy requirement, $E_0$, for erasing a single bit of information~\cite{Landauer1961}. This is given by 

\begin{align}
	E_0 = k_B T \ln(2),
\end{align}
where $k_B$ is Boltzmann's constant and $T$ is the temperature. Assuming room temperature operation ($T=293$\,K), the minimum energy required to erase a single bit of information is 0.0175\,eV or $2.8\times 10^{-21}$\,J. Figure~\ref{figure:Landauer} shows the Landauer limit as a plot of energy against temperature. Berut et al. experimentally verified the classical Landauer limit in 2012~\cite{Berut2012} and Yan et al. demonstrated a quantum Landauer limit in 2018~\cite{Yan2018}.

\subsubsection{The Analog Landauer Limit}
Because the Landauer limit applies to a discrete system such as a binary system of bits, one could argue that this principle does not apply to analog computations. However, a truly analog computer would require infinite precision, which is not possible, as Diamantini et al. argue~\cite{Diamantini2016}. They present an analog generalisation for the Landauer limit, given by:

\begin{align}
	-\frac{(\Delta) (S)}{k_B N_L} = \ln \left( \frac{8\pi s_{\mathrm{clas}}}{\hbar} \right).
\end{align}
Here, $\Delta$ is a factor, $p_0(M)$, which gives the density of the discrete probability distribution, $p(x)$, that represents the spin degree(s) of freedom, i.e., the minimum quantum of the configuration volume of a physical system. 
$S$ is the Shannon entropy, $k_B$ is the Boltzmann constant, $\hbar$ is Planck's constant, and $s_{\mathrm{clas}}$ is the electron spin in the classical limit of the quantum Heisenberg model. $N_L$ is the integer ($\mathrm{int}$) number of spins, given by:

\begin{align}
	N_L = \mathrm{int} \left(\frac{8\pi L}{\hbar} \right),
\end{align}
where $L$ is the angular momentum of the system. This allows for the number of bits, $n$, which characterise the system, to be determined from $n = \log_2(N_L)$. For the Heisenberg model, one can assume that $s_\mathrm{clas} = 1$ , which reduces the expression to $\ln(\frac{8\pi}{\hbar})$.

Landauer derived the original limit in the nascent days of statistical thermodynamics, applying it only to equilibrium, non-reversible digital systems, i.e., a Boolean logic gate. For conventional computing, this makes it a useful guideline. More recent work on the link between logical and thermodynamic reversibility has shown that the thermodynamically reversible erasure of a bit is not forbidden by the laws of physics~\cite{Sagawa2014}, however irreversibility may be an inherent aspect of a computation.

\subsection{Case Study: Optical Neural Networks Below the Landauer Limit}
We have seen that there exists a fundamental, thermodynamic limit on information processing, namely the erasure of a single bit. Hamerly et al.~\cite{Hamerly2019} detailed an optical neural network which, theoretically, had a lower bound on a single multiply-accumulate (MAC) operation below the classical Landauer limit.

The authors describe the scenario of a MAC operation using 32-bit numbers. This would require on the order of 10\textsuperscript{3} binary logic gates, which are subject to the original Landauer limit as they are digital and irreversible. This has a corresponding lower bound energy per MAC operation ($E_\mathrm{MAC}$) of $E_\mathrm{MAC}\geq 3$\,aJ. In contrast, they claim a theoretical lower bound of 50\,zJ/MAC for their method, limited by shot noise in their photodetectors, with a range of 50\,zJ/MAC-5\,aJ/MAC. The authors note the consideration of the electrical energy required to generate weights for the network, estimating a minimum of 1\,fJ/MAC, but also state that existing technology should permit sub-fJ/MAC operation.

How then does the authors' system bypass this limit? The first way is the nature of the processing. It is analog and so will be subject to a different limit than the classical limit. Using Equation (8), we can estimate an analog limit. Continuing with the $n=32$ bit example, $N_L=2^{32}$. The authors use a detector with one million pixels (i.e., a 1000$\times$1000 pixel sensor), where each pixel is 20\,$\mu$m $\times$ 20\,$\mu$m, but confine the active area to a 5\,$\mu$m $\times$ 5\,$\mu$m space. Analog signals become discretised at the point of measurement and so we will consider the detector as the computing medium here. Let us assume the thickness of each pixel is 100\,nm and that the average diameter of each atom is 0.3\,\AA. $\Delta$ is in terms of atoms and so each dimension of the pixel is divided by the atom diameter to gain the value in terms of atoms. To calculate the energy, we rearrange Equation (8) and multiply by temperature, $T$:%\todo{The maths will need verifying, I'm not completely confident that I've understood/interpreted all of the terms in Diamantini's work correctly.} 

\begin{align*}
	E = TS =  -\frac{k_B N T}{\Delta} \ln\left(\frac{8\pi s_{\mathrm{clas}}}{\hbar}\right).
\end{align*}
Assuming room temperature operation, the per-pixel energy requirement ($E$) of about $6\times10^{-25}$\,J and a total of $6\times10^{-18}$\,J for all one million pixels, which falls within the ballpark of the authors' claims. We have made a number of assumptions based on the available information so our calculations likely deviate from the true numbers. For comparison, Yan et al. give a quantum Landauer limit of $1.7\times10^{-28}$J at 48.5\,$\mu$K and so Hamerly et al.'s lower bound does not appear to violate any of these limits.

The other aspect of Hamerly et al.'s work that exempts it from the classical Landauer limit is the fundamental operation is reversible. The dot product computation is computed by the homodyne detector using a polarisation identity which uses optical interference, a reversible process, to convert the signals representing the vectors into the correct form. Here then we have a good example of how, by harnessing specific properties of a material or medium, one can make significant improvements in computational speed and efficiency.

\subsection{Case Study: Memcomputing}
\label{section:memcomputing}
Complexity is an essential consideration when it comes to computing approaches, but it is often overlooked by the devices community. A good example of this is the paradigm of \textit{memcomputing}. Amongst the devices community, it appears to be an elegant and promising solution to the \textit{von Neumann bottleneck}, which is framed as a problem of moving data. The underlying premise is that, rather that moving data between a processor and a memory element, the processing and memory elements are combined into a single unit and so processing is done in-memory~\cite{Ventra2013}. 

This begs the question: why has the paradigm been met with such a lukewarm reception by computer architectures researchers? One claim in particular, that it is possible to efficiently solve $NP$-complete problems in polynomial time using memcomputing~\cite{Traversa2015}, has, understandably, led to doubt by the architectures community. This is, in practise, a claim to answer one of the major unsolved problems in computer science, the \textit{$P$ vs $NP$ problem}~\cite{Cook2000}. A simple formulation of the problem is as follows: \textit{can a problem whose solution can be verified quickly also be solved quickly}? In other words, can a problem whose solution can be verified in polynomial time also be (efficiently) solved in polynomial time? The general consensus amongst the architectures community is the $P\neq NP$, whereas the claim of memcomputing implies that $P=NP$. Such a discovery would have a significant impact across a multitude of fields.

Traversa et al.~\cite{Traversa2015} present a \textit{memprocessor} consisting of four fundamental computing elements: a differentiator, four voltage multipliers, a summing amplifier and a difference amplifier. The memprocessor encodes information in the frequencies of the collective state of signals in the machine, i.e., the measured signal is the sum of the encoding frequencies. The system does readout by taking the Fourier transform of the signal. For a demonstration of an $NP$-hard problem, they use a version of the \textit{Subset-sum problem}. The full subset-sum problem asks the following question:
for a set of $N$ integers, $k$, is there a subset, $s$ which sums to a particular number, $P$? A common formulation of this has $P$ as zero and $s$ as a non-empty subset, i.e., if $k = \lbrace -2, 4, -7, 9, 10 \rbrace$, then we can see that there exists a subset $s = \lbrace -2, -7, 9 \rbrace$ which will sum to zero. As the \textit{power set}, i.e., the set of all subsets contains $2^N$ subsets, then it is clear that the problem size grows exponentially with the input size. Indeed, algorithms to solve this problem have exponential time complexity, with the naive solution having complexity $\mathcal{O}(2^N N)$ and better approaches having complexity $\mathcal{O}(2^{\frac{N}{2}})$. 

Traversa et al. generate waves at frequencies that encode all possible subsets of $k$ and measure the output signal to see if a frequency corresponding to $P$ exists. As the number of subsets grows exponentially with input size, then, despite anything else, exponentially many frequencies will be required to encode all possible subsets. This measurement therefore will therefore take an exponential amount of time. If each measurement takes the same amount of time, an exponentially increasing number of measurements will take exponentially increasing time. 

The authors also remark that ``all frequencies involved in the collective state... are dampened by the factor $2^{-N}$'', in order to place a bound on the energy consumption of the system. This also presents a fundamental limit of the scalability of the system, beyond the obvious issue of exponential growth when taking into account the Landauer limit. From the authors' work, we see that energy is related to the fundamental frequency, $f_0$ by $E \leq \frac{1}{f_0}$, i.e., the energy for a computation is proportional to the energy in the collective state in a single period.  Frequency is related to wavelength $\lambda$ by the well-known relation $f=\frac{c_0}{\lambda}$, where $c_0$ is the speed of light. We can establish the smallest fundamental frequency by setting the Planck length as the shortest possible wavelength. From this, we can get the following expression for the maximum energy of the collected state (the sum of frequencies):

\begin{align*}
	E \leq \frac{\lambda}{c_0}.
\end{align*}
As we know that the attenuation is by a factor of $2^{-N}$, we have an expression for the maximum energy of the collected state $E$ and a thermodynamic limit on the energy for processing a bit of information, $E_0$, we can combine these into an expression to find the maximum input size:

\begin{align*}
	2^{-N} E \geq E_0.
\end{align*}
Rearranging for $N$:

\begin{align*}
	N \leq \log_2 \left( \frac{\lambda}{c_0 E_0} \right).
\end{align*}
From this, we can finally establish a maximum value of $N$ based on fundamental physical constants, which turns out to be about $N \leq 76$. Again, this is proportional to the energy in the collective state.

This does not dismiss the work on memcomputing, because the authors do indeed demonstrate a speedup and improvement over existing methods for the problem, verifying this in more recent simulations~\cite{Traversa2018}, albeit for approximations rather than true solutions. The issue is the exponential factor. Though the memprocessor solves a particular problem in a reasonable time, higher calculation precision requires exponentially greater measurement precision, meaning there is a trade-off between these, and scalability becomes a limiting factor.

\subsection{The Unavoidability of Computational Complexity}
We have seen that memcomputing is still subject to the same constraints of computational complexity as any other architecture. Are there other architectures that might escape this, such as quantum computing, which can try every solution to a problem at once? The answer, unfortunately, is no. Despite their speed and special properties, quantum computers are still subject to the same constraints of computational complexity. On a basic level, this is because quantum mechanics can, to the best of our understanding, describe all physical phenomena. This means that any problem solvable by a classical computer can also be solved by a quantum computer and vice-versa. 

Aaronson~\cite{Aaronson2005} addresses the issue of exponentially growing resources, evaluates various proposals for solving NP-complete problems in polynomial time under known physics, and suggests that efficient solutions to NP-complete problems are unlikely and that such attempts are akin to attempts at constructing perpetual motion machines. This is a powerful guiding principle in development of computing devices, as, like the various forms of the Landauer principle, it helps place constraints on possible devices and computations.

The scalability consideration also suggests a limit of what researchers in the field should concern themselves with; a solution to the $P$ vs $NP$ problem is the domain of theoretical computer science and mathematics. Attempting a proof via demonstration is akin to demonstrating infinity by writing out all its digits.

One final point worth reiterating is the nature of exponential growth. A common argument is that smaller devices and increased integration density will offset the problems of exponential growth in the number of devices required by an approach, yet this does not hold up to scrutiny. Consider some hypothetical computing device of dimensions 1$\times$1$\times$1\,nm that can perform any desirable computation on a single bit of information. Say an arbitrary number of these are integrated together into an array on a single die. Now say we wish to implement an algorithm of circuit complexity $\mathcal{O}(2^n)$ in hardware on this system. If we start with some operation involving two 4-bit numbers, then the system would require $2^4=16$ devices to function, giving a total lateral area of 16\,nm\textsuperscript{2}. If we then jump to a typical binary, single-bit floating point representation, i.e., 32-bits, this requires $2^{32}=4,294,967,296$ devices, giving a lateral area of about $4\times10^9$\,nm\textsuperscript{2} for two numbers. If we wished to use such a system for a cryptographic computation, using a standard 128-bit encryption, this would require $3.4\times10^{38}$ devices, corresponding to a lateral area of $3.4\times10^{20}$\,m\textsuperscript{2}. For comparison, the average surface area of the sun is about $6.1\times10^{18}$\,m\textsuperscript{2}. Even if these devices were integrated in three dimensions, the total volume would be at least $3.4\times10^{11}$\,m\textsuperscript{3}, which is about a third of the volume of Mars' moon, Deimos (volume $\approx 10^{12}$).
\section{A Question of Architectures}
\label{section:architectures}
In its simplest definition, an algorithm is a sequence of operations to perform a computation or calculation, independent of implementation. A computer program is an implementation of an algorithm. Likewise, a hardware accelerator, such as a ReRAM crossbar for matrix-vector multiplication (MVM), is also an implementation of an algorithm. 

\textit{Architectures} is also a nuanced term. In computing, there are two main things which the word \textit{architecture} refers to. The first is an \textit{instruction set architecture} (ISA), which is the abstract model of a computer and is what the term \textit{computer architecture} is typically used to refer to. An ISA describes the data types, registers, I/O model and fundamental features of a computer. The two main classes of ISA are Complex and Reduced Instruction Set Computer (CISC and RISC) architectures. The Intel x86 family is an example of a CISC architecture and the ARM family is an example of a RISC architecture.

Like an algorithm, an ISA can be implemented in different ways. For example, though AMD and Intel might both build CPUs with an x86 ISA, the physical implementations will be very different. This is an example of the other use of the term \textit{architecture} that commonly arises in discussions of computer systems, referring here to a \textit{microarchitecture}, which is the set of processor design techniques used in a given piece of hardware, i.e., the hardware layout.

\begin{figure}[h]
	\centering	
	\includegraphics[width=\linewidth]{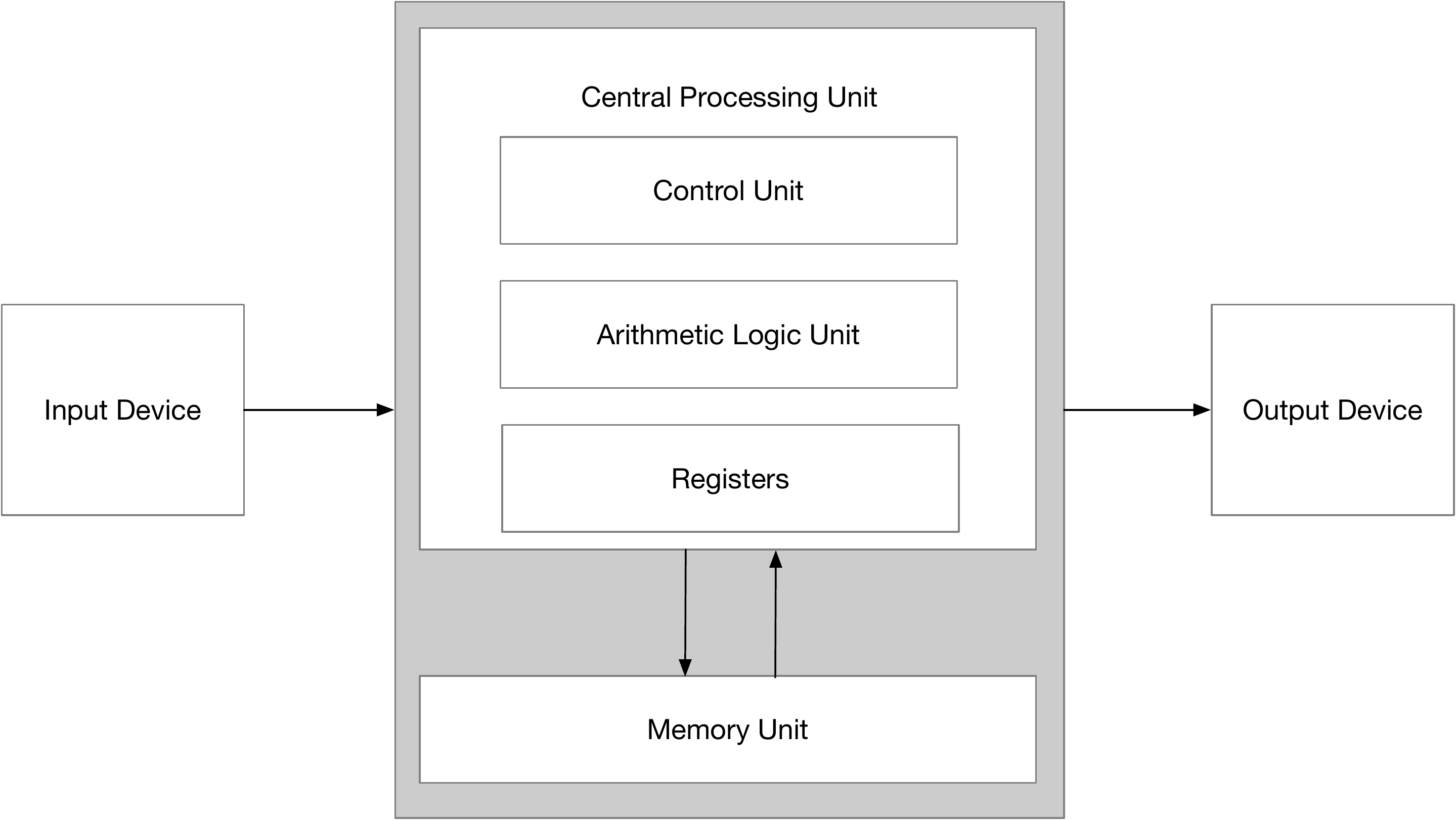}
	\caption{Schematic of a von Neumann computer.}
	\label{figure:vonNeumann}
\end{figure}

\begin{figure}[h]
	\centering
	\includegraphics[width=\linewidth]{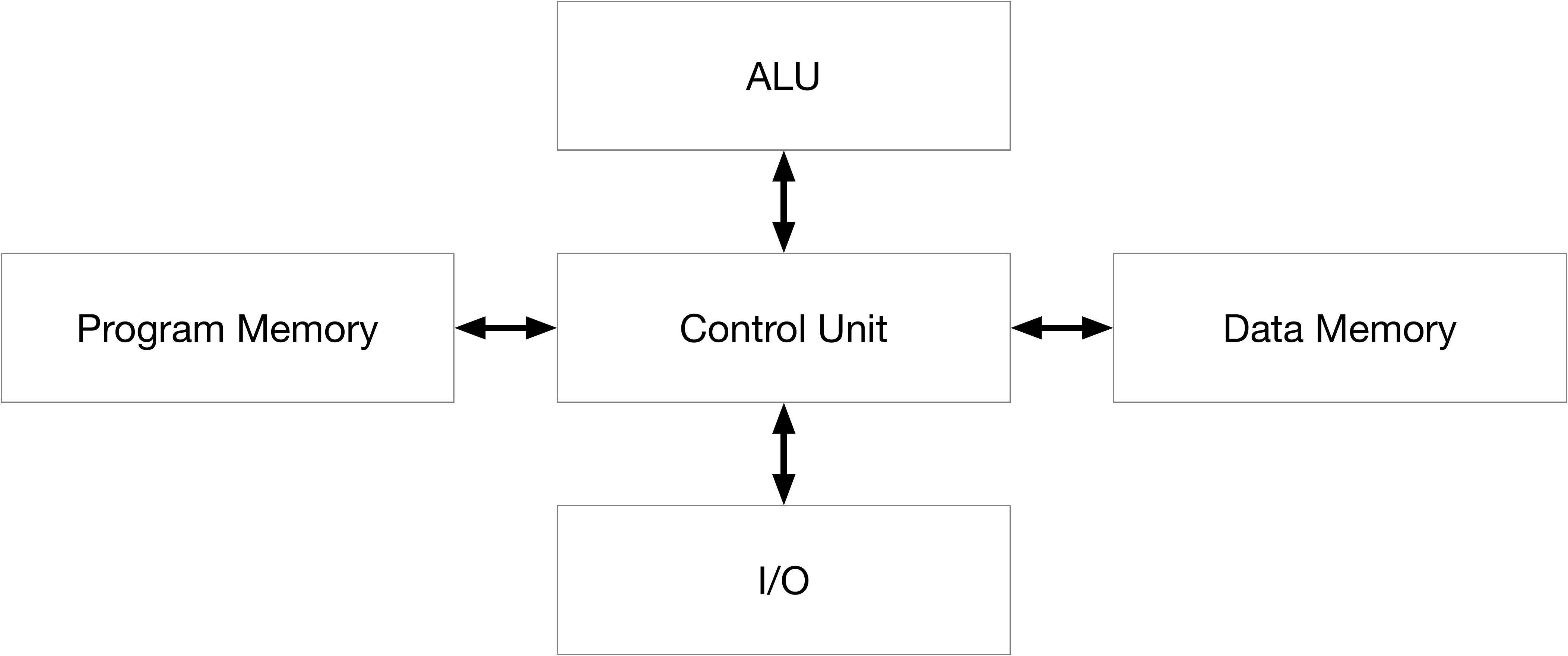}
	\caption{Schematic of the Harvard architecture.}
	\label{figure:Harvard}
\end{figure}

Most modern computers are based on either the \textit{von Neumann} architecture (also known as the \textit{Princeton} architecture, Figure~\ref{figure:vonNeumann}) of the \textit{Harvard} architecture (Figure~\ref{figure:Harvard}). The former was developed by von Neumann in 1944~\cite{vonNeumann}, itself inspired by the \textit{universal computing machine} described by Turing in 1936~\cite{Turing1936}.

However, the von Neumann architecture has a fundamental design limitation which hinders performance, the \textit{von Neumann bottleneck}, which arises from the shared bus used by the program and data memory. This limits the throughput between the CPU and the memory, as only program or data memory can be accessed at a given time and so the CPU must wait for data to be moved to and from memory to process it. No matter how much CPU speed and memory size increase, the limited throughput means that there will always be a significant restriction on data processing rates.

The Harvard architecture does not use a shared memory and program bus and so is not subject to the associated bottleneck. Though this avoids the von Neumann bottleneck, it also means that the data and program memory in a Harvard computer occupy different address spaces. Here, a memory address does not identify a storage location and so the memory space that the address belongs to, whether data or program, must also be known. This gives von Neumann computers an advantage in the sense that data and program memory can be treated in the same way. For example, data can be read and then executed as code, whereas a Harvard machine requires additional processing.

The Harvard architecture is commonly used in digital signal processors~\cite{Tan2019} as well as a number of microprocessors including the ARM Cortex M3~\cite{Yiu2010} and PIC microcontrollers~\cite{Wolf2017}. Many modern general computers use a \textit{modified Harvard architecture}, which is similar to the von Neumann architecture in that there are separate CPU caches for data and instructions which share a single address space. This is the split-cache architecture of most modern CPUs. 

An architecture can have multiple implementations. Many research articles which propose and demonstrate novel devices make reference in some form to the limitations of contemporary computer architectures, before going on to describe the implementation of a particular algorithm, e.g., a MVM accelerator/dot-product engine using a novel device. This comparison of an abstract computer architecture to an algorithm implementation is thus meaningless. Consider again the von Neumann architecture. One of the components of this is the ALU. This is a digital circuit which performs arithmetic and bitwise operations on integer binary numbers, i.e., a circuit which implements algorithms (binary addition, multiplication, division etc.). Likewise, a circuit which implements a MVM operation is simply that. It is not a computer architecture in and of itself, but may be used in the hardware microarchitecture. 

%Consider the use of memristors for acceleration of machine learning inference. Due to the nature of the crossbar approach, a memristor-based crossbar accelerator reduces the complexity of the MVM operation from  $\mathcal{O}(n^2)$ to $\mathcal{O}(1)$~\cite{Hu2016}. We elaborate on the significance of this in Sections~\ref{section:complexity} and~\ref{section:ml-accelerators}. We can see then that novel devices and circuits can offer considerable improvements over existing methods of computation, however it is important to frame and contextualise the research being done in order to maximise its impact.
\section{Brief Overview of Machine Learning}
\label{section:ml}
Machine learning is a field which has, in recent years, become simultaneously a technological revolution, a household term, a source of moral and philosophical consternation, and a scientific endeavour. Despite its long history, with the term \textit{machine learning} being coined in 1959~\cite{Samuel1959}, it is only relatively recently that ML has entered mainstream awareness. ML has quickly become a dominant technology, seeing applications in almost all areas of computing, from computer vision~\cite{Sebe2005} to medicine~\cite{MLMedicine} and law~\cite{MLLaw}.

At a fundamental level, ML is where a program is able to make predictions from a pre-existing dataset. In strictly mathematical terms, a dataset with one or more features (variables) has an associated output. ML is the attempt to discover the mathematical function that depends on these features and most accurately predicts the output, though ML is a broad term which encompasses a wide range of approaches and models. Domingos~\cite{Domingos2012}, suggests that all learning algorithms have three principal components: \textit{representation}, \textit{evaluation}, and \textit{optimisation}. Commonly-used terms, such as ``machine learning techniques'' typically refer to a given representation, e.g., neural networks, independent of the evaluation function or optimisation methods. 

\subsection{Representation}
Domingos~\cite{Domingos2012} gives six classes of representation for machine learning methods: \textit{instances} (K-nearest neighbour, support vector machines), \textit{Hyperplanes} (naive Bayes, logistic regression), \textit{Decision Trees}, \textit{Sets of Rules} (Propositional rules, logic programs), \textit{Neural Networks} and \textit{Graphical Models} (Bayesian networks, conditional random fields).

\subsubsection{Neural Networks}
\label{section:NNs}
\begin{figure}[h]
	\centering
	\subfigure[]{\includegraphics[width=0.9\linewidth]{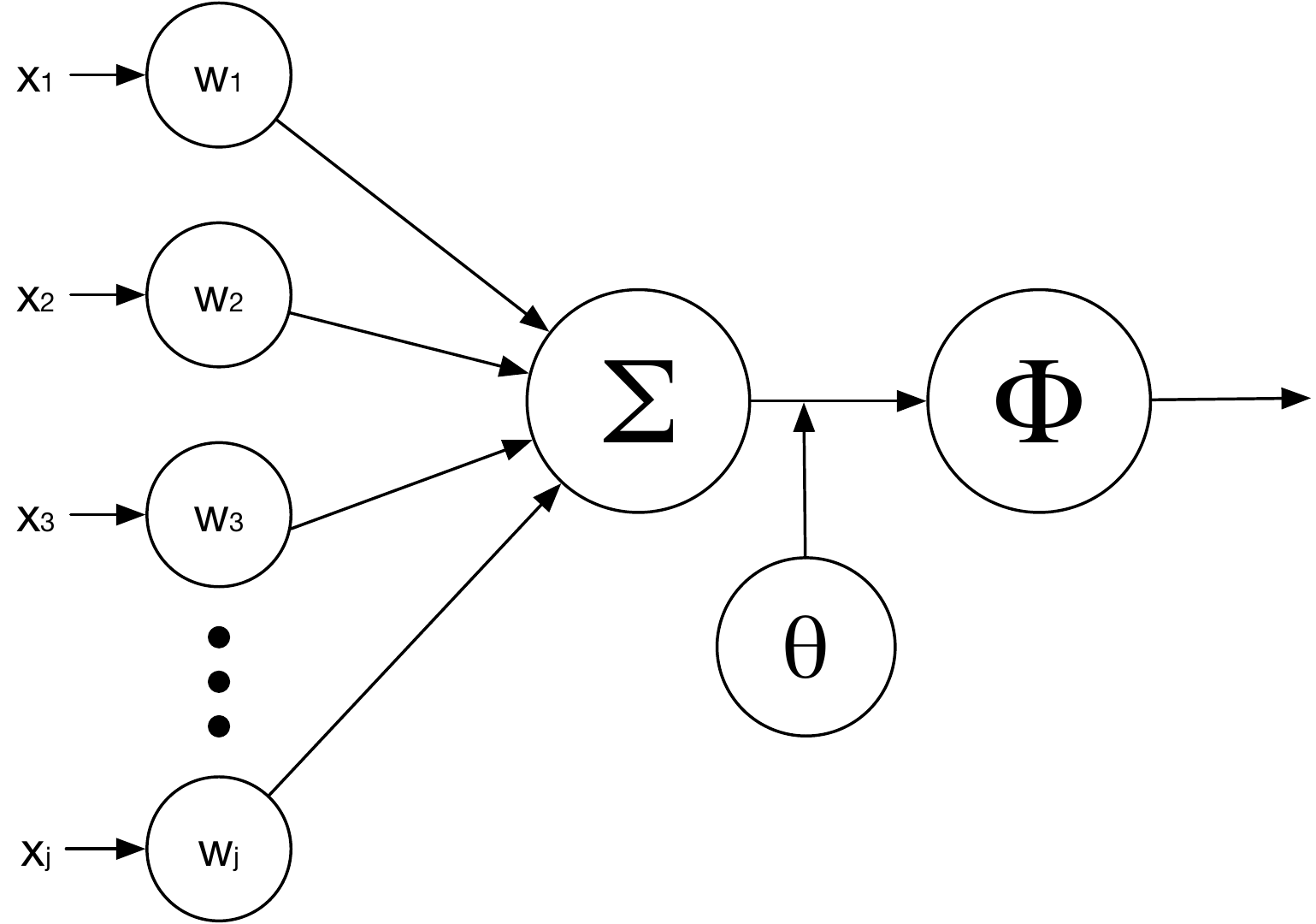}}
	\subfigure[]{\includegraphics[width=0.9\linewidth]{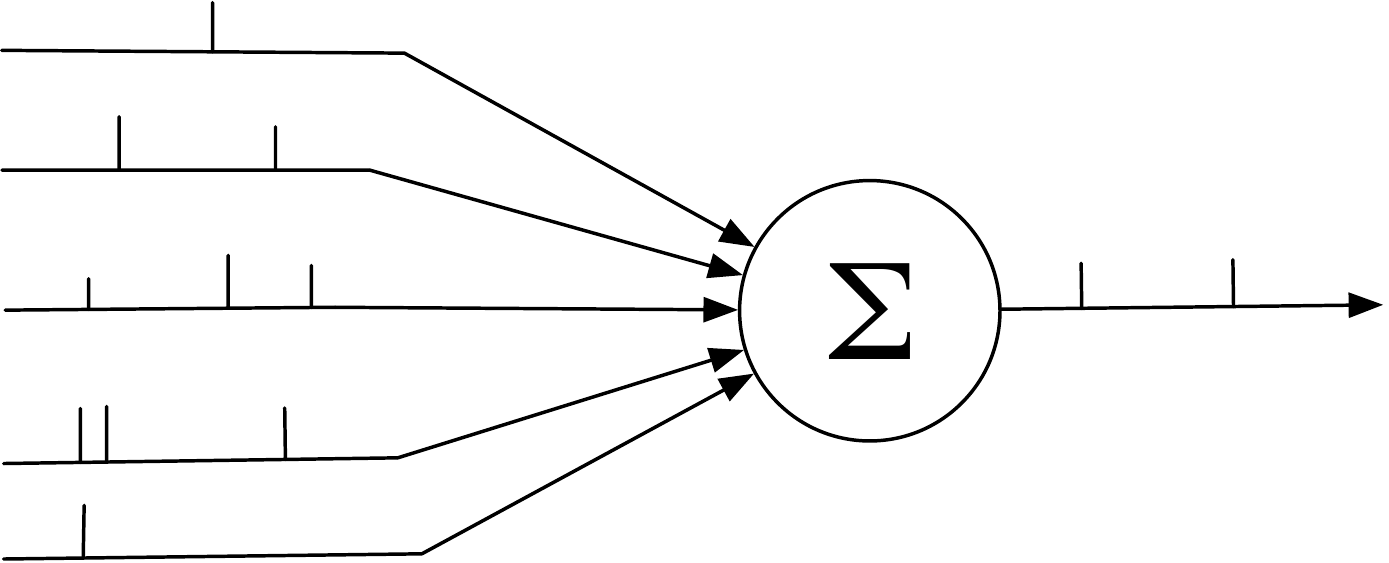}}
	\caption{\textbf{(a)} Schematic of a simple single-layer perceptron ANN; \textbf{(b)} Schematic of a simple, single-layer spiking neural network}
	\label{figure:ANNSchem}
\end{figure}

ANNs are the most ubiquitous ML approach and are modelled on biological neurons. The simplest types of neural network are a type of feedforward neural network known as single-layer perceptrons. These consist of an input layer and a single output neuron, corresponding to a binary output. More complicated ANNs will have one or more hidden layers and an output layer. During training, each input will fire and activate a neuron in the next layer and different inputs will fire with different intensities (weights). Figure \ref{figure:ANNSchem} \textbf{(a)} illustrates a simple ANN with a set of inputs ($x_{1}$, $x_{2}$, $x_{3}$, ... , $x_{j}$) and corresponding weights ($w_{1}$, $w_{2}$, $w_{3}$, ... , $w_{j}$), a transfer function ($\Sigma$), an activation function ($\Phi$) and a threshold ($\Theta$).

The error is calculated using a cost function such that $E = L(t,y)$. A commonly-used function is the squared error, $(t-y)^2$, where $t$ is the true value and $y = x_1w_1 + x_2w_2 + x_3w_3 + ... + x_jw_j$. In each training cycle, the weights of the connections between neurons in each layer are adjusted until the weights giving the most accurate classification of the test sets are determined. Optimal weights are found by making use of the cost function, which measures the discrepancy between the neural network's classification and the actual result. This updating of weights is the most vital aspect of training and is known as \textit{backpropagation}.

Feedforward networks can be extended beyond the simple single-layer perceptron example, and most state-of-the-art feedforward networks involve multiple layers between the input and output. Hence, they are described as \textit{deep} neural networks (DNNs). Depending on the application, different varieties of DNN are used. Recurrent neural networks (RNNs) are common for applications with sequential data, such as speech recognition, whereas convolutional neural networks (CNNs) are the predominant approach for tasks such computer vision, natural language processing and image classification.

Spiking neural networks (SNNs) differ from feedforward networks in that each neuron only fires once a specific threshold value is reached, thus SNNs more closely resemble biological NNs than feedforward networks. Figure~\ref{figure:ANNSchem} \textbf{(b)} The threshold for a neuron firing is commonly referred to as the \textit{membrane potential}, after the property in biological cells. These have an inbuilt time dependence, with input taking the form of a spike train rather than an instantaneous value. This means that, in a given epoch (training cycle), not all neurons will activate, and that a given neuron requires a certain time to activate, i.e., when a sufficient number of spikes have been input within a given time.

\subsubsection{Instances}
$k$-nearest neighbours ($k$-NN) is a method used for both classification and regression. In the former case, the algorithm output is the class to which the object being tested belongs to. In the latter, the output is the property value of the object of interest. A $k$-NN algorithm classifies an object by measuring the distance between a datapoint of interest and neighbouring datapoints with the final classification of the datapoint determined by the class of the $k$ nearest neighbours, where $k$ is a small, positive integer value.

Support Vector Machines (SVMs) are ML models which are used to find and construct hyperplanes.
A hyperplane is defined geometrically as a space with one less dimension than the ambient space. For example, a hyperplane would be a 2D plane in 3D space. In the case of ML, these act as boundaries between classes of datapoints. Datapoints that fall on one side of a boundary can therefore be considered as belonging to one class, whereas those that fall on another side belong to a different class. A support vector is a datapoint which is close to the hyperplane and helps influence the hyperlane's position and direction. These support vectors can be used to find the margin of a classifier, which is the distance from a datapoint to a decision boundary. Ideally, this will be large.

\subsubsection{Hyperplanes}
For SVMs to be able to draw the boundaries between classes of data, algorithms which classify the data to begin with are needed. Na\"{i}ve Bayes methods, such as Gaussian or Multinomial Na\"{i}ve Bayes classifiers, are an example of these. Na\"{i}ve Bayes refers to a family of classifiers which use Bayes theorem, given as follows:

\begin{equation}
P(B|E) = \frac{P(E|B) P(B)}{P(E)}.
\end{equation}
Here, $B$ and $E$ are two independent random variables. For the example here, $B$ is a belief and $E$ is a piece of evidence. Bayes' theorem states that the probability a belief is true, given a new piece of evidence, is equal to the probability that evidence is true given that the belief is true, multiplied by the probability that the belief is true, regardless of the new evidence, all divided by the probability that the evidence is true regardless of whether the belief is true. Na\"{i}ve Bayes algorithms are so-called because they assume a strong independence between features in a dataset.

\subsubsection{Decision Trees}
Decision tree (DT) learning is a predictive model approach which uses decision trees to draw conclusions about the value of some item, based on observations about the item. A DT represents observations the branches of the DT and the conclusions in the leaves. DTs where the variables have discrete values are known as classification trees, whereas those where the variables are continuous are known as regression trees.

\subsubsection{Graphical Models}
Bayesian inference is another approach based on Bayes' theorem. Bayesian networks can be considered probabilistic computers, as they compute probabilities as opposed to specific numbers or values. 

Conditional Random Fields (CRFs) are a method which differ from classifiers such as Naive Bayes or $k$-NN in that CRFs consider context, whereas the others do not. Random fields themselves are a function that takes on a random value at a given point in the input space, which is often multi-dimensional. A given CRF models the conditional probability distribution of a graph whose nodes form two disjoint sets, one of the observed variables and one of the output variables.

\subsection{The Curse of Dimensionality}
Real-world datasets are often very large, with many variables, which gives rise to the `curse of dimensionality', coined by Bellman in 1961~\cite{Bellman1961}. This refers to the phenomena arising from analysis and organisation of high-dimensional datasets, namely the tendency for algorithms working well in lower dimensions to become far more complex as the dimensionality increases. For ML, this is a significant issue, as data becomes sparser and meaningful groupings or methods that rely on statistical significance break down, something which Domingos~\cite{Domingos2012} addresses.

\subsection{Machine Learning Benchmarks}
With a range of ML approaches being actively researched, it is vital that some common frame of reference or benchmark exists which allows for different methods to be evaluated against one another in a meaningful way. Here we present a brief overview of some popular ones.

\subsubsection{MNIST}
The most common benchmark, which serves as the ML analogue of ``Hello World'' in traditional computer programming, is the MNIST dataset. This is an optical character recognition (OCR) task using a database consisting of 60,000 training examples and 10,000 test examples of handwritten digits, where each example is a 28$\times$28 pixel greyscale image. The current top five verified algorithms for MNIST have accuracies of around 99.79\,\% or greater~\cite{MNISTLeaderboard}, whereas accuracies closer to 90\,\% are often reported for experimental devices such as P-bits~\cite{Pagliarini2020} or ReRAMs~\cite{Hu2018}. Such accuracies are significantly lower than those of MNIST software approaches from as far back as 1998~\cite{Lecun1998}.

\subsubsection{UCI Databases}
Another popular and comprehensive suite of benchmarks are the 559 (at the time of writing) datasets provided by researchers at the University of California, Irvine~\cite{Dua2019}. These datasets cover a broad range of fields and data types (e.g., text, images, numerical), from healthcare (e.g., arrhythmia, lung cancer, thyroid disease), botany (types of iris plant, types of soybean), finance (credit approval, stock exchanges), and many others. As ML algorithms are typically tailored to the data from which one wishes to learn, these datasets offer an excellent resource for researchers wishing to test their algorithms.

\subsubsection{ImageNet}
ImageNet is another popular benchmark, consisting of over 14 million images and focuses on image classification~\cite{Deng2009}. Unlike MNIST, which covers just handwritten characters, ImageNet includes datasets of fungi, flora, fauna, geological formations, people, and miscellaneous objects such as food items.

\section{Novel Devices as Alternatives to Digital Silicon Electronics}
\label{section:computing-devices}

\subsection{Limitations of Digital CMOS}
The key driver in increasing computer performance over the past several decades is the miniaturisation of computing devices. Moore's law, which described the apparent doubling in the transistor count of integrated circuits (ICs) every eighteen months~\cite{Moore1965} became a guiding principle in the development of semiconductor roadmaps. Another law which guided the development of semiconductor devices, and CPUs in particular, for around thirty years is \textit{Dennard scaling}~\cite{Dennard1974}. Dennard et al.'s article describes scaling relationships in MOSFETs, providing a framework within which device designers can scale various parameters to improve device performance. The key result of the article is constant field scaling, which describes the relationship between the supply voltage ($V_\mathrm{dd}$) and dynamic switching power ($P_\mathrm{dyn}$) via the relation $P_\mathrm{dyn} \propto C V_\mathrm{dd}^2$, where $C$ is the gate capacitance. 

\begin{figure}
	\centering
	\hspace*{-0.3cm}\includegraphics[trim= 0cm 2.1cm 0cm 0cm, clip=true, width=1.035\linewidth]{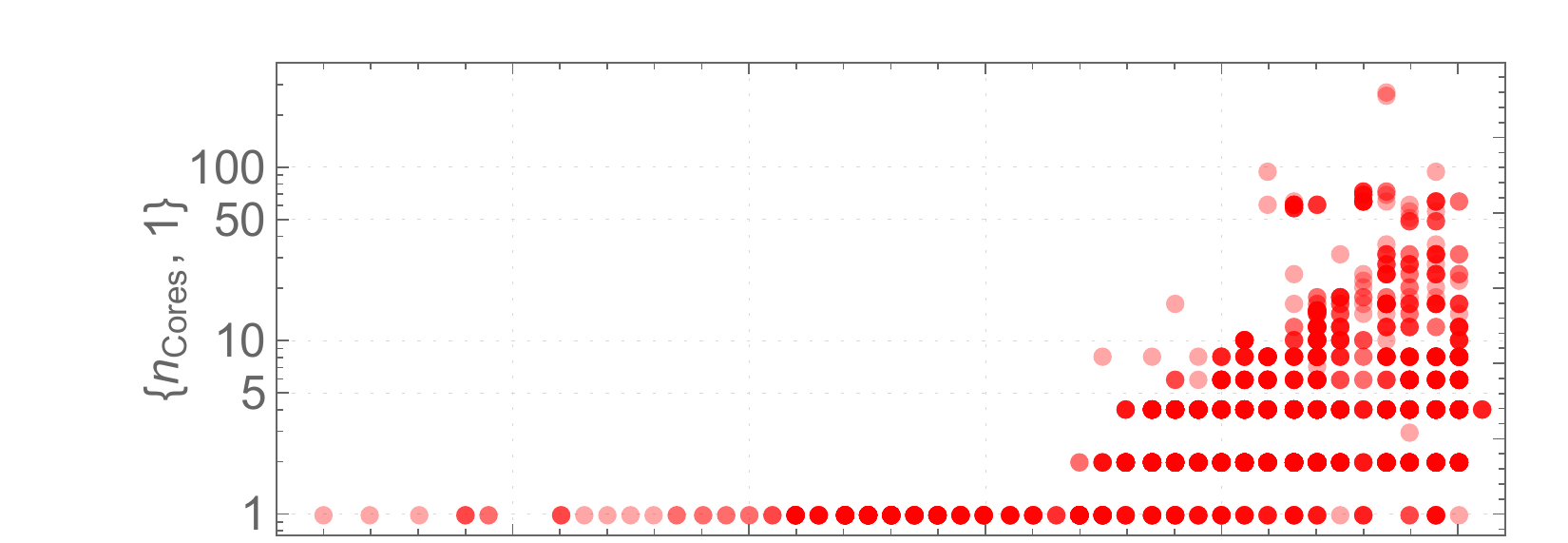}\llap{\makebox[7.3cm][l]{\raisebox{2.4cm}{\textbf{(a)}}}}\vspace{-0.168in}
	\hspace*{-0.55cm}\includegraphics[trim= 0cm 2.1cm 0cm 0cm, clip=true,width=1.065\linewidth]{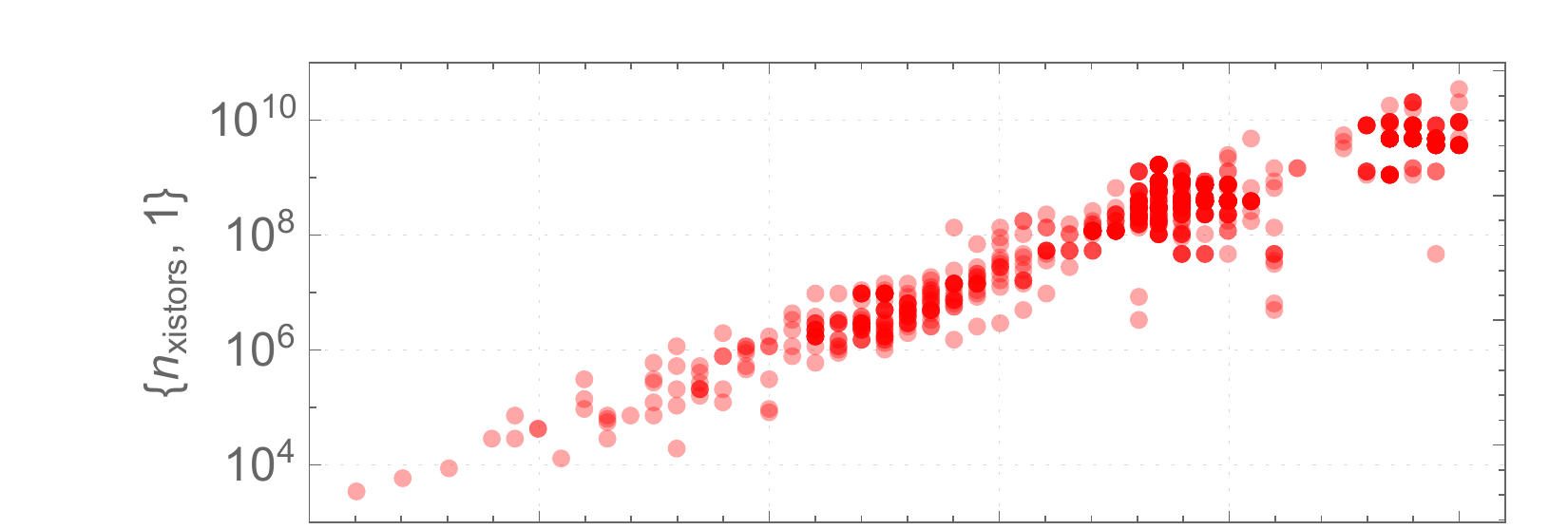}\llap{\makebox[7.3cm][l]{\raisebox{2.4cm}{\textbf{(b)}}}}\vspace{-0.1472in}
	\includegraphics[width=\linewidth]{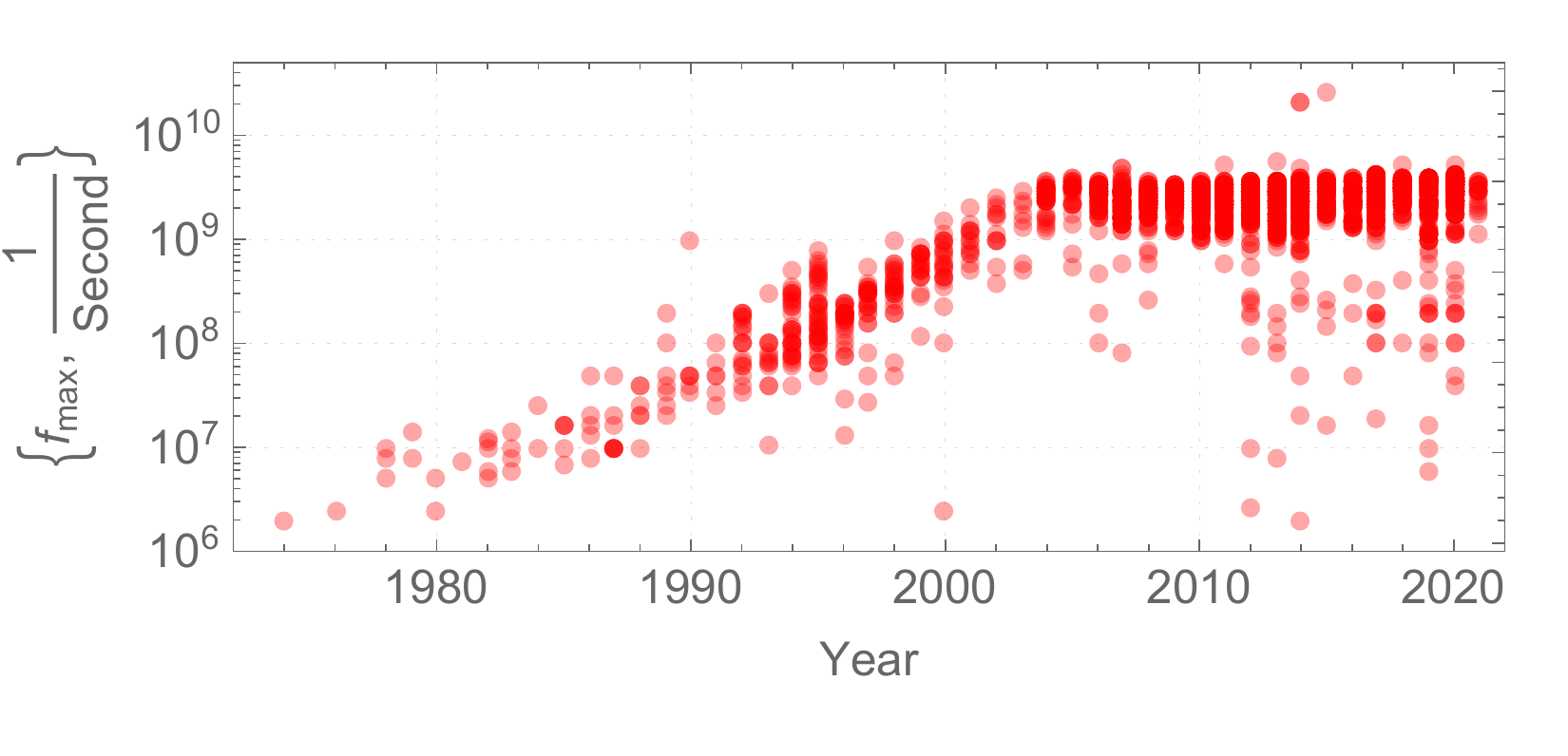}\llap{\makebox[7.3cm][l]{\raisebox{3.5cm}{\textbf{(c)}}}}\vspace{-0.172in}
	\vspace{-0.5cm}
	\caption{Plots of various trends in integrated circuits over time. \textbf{(a)}: number of CPU cores over time; \textbf{(b)}: CPU transistor count over time; \textbf{(c)}: CPU switching speed over time. The shade of the points corresponds to the number of datapoints, i.e., darker points have more data, whereas lighter ones have less.}
	\vspace{-0.5cm}
	\label{figure:CPUTimes}
\end{figure}

In 2006, the Dennard scaling era drew to a close and the clock speeds of CPU cores began to saturate as limits of power density were reached. This led to the introduction of multi-core CPUs, in which parallel processing threads are used. This tendency towards parallelisation has helped maintain Moore's law for CPUs by increasing the number or processing cores on a single CPU die, as opposed to simply increasing the number of transistors in a single processing core. Figure~\ref{figure:CPUTimes} illustrates these scaling laws over time, based on data we collected of both commercial and research devices~\cite{HistoricalCPUData}. Figure~\ref{figure:CPUTimes}\textbf{(c)} shows the growth in time of CPU clock speeds and \textbf{(a)} shows the number of CPU cores as a function of time.  The causes of this are primarily material; in the case of Dennard scaling, CMOS dynamic (i.e., switching) power consumption is proportional to the frequency and so reducing the size of transistors allows for faster switching. However, as devices reach a certain scale, typically below 45\,nm, it is no longer the switching which is the main source of power consumption, but leakage current~\cite{Kim2003}. Figure~\ref{figure:CPUTimes}\textbf{(b)} shows the slowing of Moore's law since around 2010.

Keyes~\cite{Keyes1985} highlights the successes of CMOS and outlines the general features that a digital computing device must have. These are: 

\begin{enumerate}
	\item Gain: High gain in each component reduces issues arising from inter-device variation. As environmental and manufacturing variations can influence device performance, signal degradation is prevented by using reference signal levels throughout a system. Signal levels are restored to this reference at each step.
	\item Input/Output Isolation: Inputs and outputs must be isolated from one-another to ensure calculations are carried out in a predetermined manner. This means the output of a device must have no impact on the input, otherwise the state of the inputs may change and influence the output in an uncontrolled manner.
	\item Comparable On/Off Switching Times: The switching time between on and off states must also be comparable. If this is not the case, a separate reset operation is needed, adding both time and material costs. 
	\item Inversion: A computer device must be able to convert a one to a zero and vice-versa.
\end{enumerate}

We have focused so far on CPUs, however Moore's Law and Dennard Scaling also apply to other technologies, including reprogrammable hardware such as field-programmable gate arrays (FPGAs), and memory such as DRAM, SRAM and Flash. The two main forms of volatile memory are SRAM and DRAM, whereas Flash memory is non-volatile. An SRAM cell typically consists of six transistors, and is typically used for the cache on a CPU, meaning it has generally scaled with CPU process nodes. However, DRAM, which requires a capacitor and a transistor, and is typically used on separate chips (e.g., the main memory in a computer). DRAM is slower but cheaper than SRAM, however has further scaling is becoming increasingly difficult, owing to the difficulties in scaling down capacitors~\cite{Park2015_2}. As we discuss in Section~\ref{section:ml-accelerators}, researchers have demonstrated ML accelerators using both DRAM and SRAM. Flash memory, which consists of a floating gate transistor, has successfully experienced 3D integration~\cite{Siau2019}, however the read/write times for it are slow, making them unsuitable for use as anything other than storage devices, as per Keyes' criteria, as well as the requirement for block-by-block erasure, as opposed to specific updating of individual bits. However, there has been some recent work in this area~\cite{Ko2020}~\cite{Lee2020}, and novel approaches may render this a feasible method.

Traditional semiconductor roadmaps predominantly rely on the continuation of scaling laws, and thus a key focus of devices and materials research is focused on replacements to conventional silicon and digital CMOS. However, many novel technologies fail to compete with digital CMOS on matters of energy, speed, scalability and price, but this does not make them non-starters for computing applications! Sarpeshkar~\cite{Sarpeshkar1998} highlights the stark difference between computational approaches, discussing the addition of two, parallel eight bit numbers which, in the analogue domain requires only a single wire, using Kirchoff’s current law. However, in the digital domain, around 240 transistors are required for a CMOS circuit. In the case of multiplication, between four and eight transistors are required for an analogue approach, whereas digital computation demands as many as 3000 transistors. It is clear that fundamentally different approaches and devices have the potential to offer significant improvements in the efficiency and resource requirements of certain computations. In practise, this is likely to mean technologies complementary to the CPU plus memory architecture of conventional computers.

\subsection{Case Study: Generating Non-Uniform Random Variates with GFETs}
Keyes' observations about what constitutes a good computer device would suggest that some experimental devices are inherently unsuitable for computation. One such device is the graphene field-effect transistor (GFET). First reported in 2007, by 
Lemme et al. in 2007~\cite{Lemme2007}, graphene FETs (GFETs) have a channel made of 
single- or multi-layer graphene, rather than a semiconducting material such as silicon or germanium. Due to their high mobilities~\cite{Chen2008} and fast switching speeds~\cite{Vicarelli2012}, GFETs attracted significant interest from researchers. However, the lack of a band gap which gives rise to
these properties also means that GFETs have low on- to off-current ratios, making 
them unsuitable for digital logic applications~\cite{Keyes1985}.

\begin{figure*}[t]
	\centering
	\includegraphics[trim= 1cm 3cm 1.5cm 1cm, clip=true, width=\linewidth]{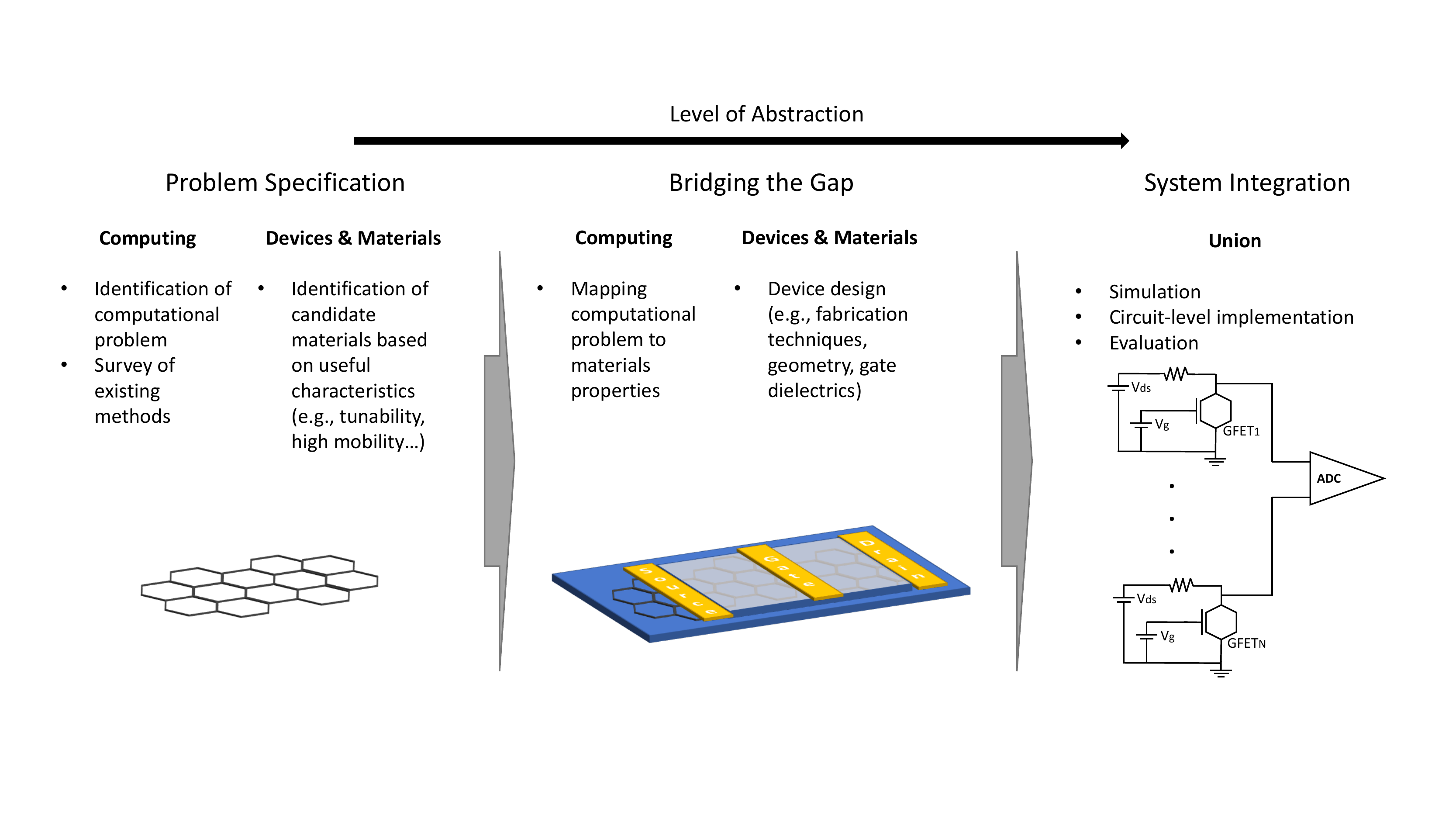}
	\caption{Overview of the development process for the GFET-based non-uniform random variate generator in~\cite{Tye2020}, showing the combination of materials and device selection as well as a mapping of a computational problem to an experimental device.}
	\label{figure:matstocircs}
\end{figure*} 

However, this does not preclude the use of GFETs for alternative computing applications. 
Our own work provides an example of this~\cite{Tye2020}. Figure~\ref{figure:matstocircs} provides an illustrated overview of the process. As a computational problem, we investigate sampling from non-uniform random variates. This is a computational problem which arises in ML applications such as Bayesian inference, which requires the computation of marginal probabilities. This involves integrating a many-dimensional probability distribution, which rarely has an analytic form. Sampling from the distribution is therefore necessary. Typically, this sampling is done using Markov Chain Monte Carlo (MCMC) methods, based on the Metropolis-Hastings algorithm~\cite{Hastings1970} or approaches which build on it~\cite{Green1995}~\cite{Liu2000}. Other methods include Gibbs sampling~\cite{Geman1984} and the Hamiltonian Hybrid Monte Carlo method~\cite{Duane1987}. These algorithms are computationally inefficient and expensive, meaning there is significant scope for improvement. 

Generally, algorithms for sampling non-uniform random variates are implemented using reprogrammable digital hardware, i.e., a programming language running on a CPU. In our approach, 
we use an array of selectively-biased GFETs to approximate a lognormal probability distribution, and predict a $\approx 2\times$ speedup compared to MATLAB, with the sample rate limited by the speed of the analog-to-digital converter (ADC). Although our initial approach produced less accurate samples than MATLAB, we demonstrate comparable levels of accuracy to MATLAB non-uniform random number generation using as few as 16 GFETs, as well as application to arbitrary probability distributions in subsequent simulations (yet to be published), by using GFET transfer characteristics as orthonormal basis functions. Being significantly faster and requiring far fewer devices than an all-digital approach, we see how, by mapping a computational problem to novel devices and materials, one can provide significant improvements over existing approaches. External circuitry is required, e.g., to calculate bias voltages, as well as ADCs to read the generated sample, however biases only need to be determined once for a given approximation. Furthermore, we also demonstrate a robustness to inter-device variation, meaning the requirement for so few devices also means that the challenge of scalability, i.e., being able to produce consistent devices at scale, becomes a non-issue. 

\subsection{Experimental Devices}

\subsubsection{Non-Silicon Transistors}
Silicon, germanium, and III-V compounds such as GaAs, InGaAs are the dominant materials used in commercial transistors, however there is a large research focus into the use of alternative materials as transistors. Carbon nanotubes (CNTs), which, depending on how they are rolled, can be semiconducting or metallic are one such material, and have been the subject of an intensive research effort as FETs~\cite{Prakash2018}, as well as having demonstrated application in ML hardware~\cite{Kim2015}.

The discovery of the electric field effect in graphene triggered an interest in atomically-thin materials, due to their inherent suppression of short-channel effects as a result of their thinness. In addition, their scalability, unique properties and potential for new phenomena arising from selective stacking of materials have also made their study attractive to device researchers. FETs have been demonstrated using a range of 2D materials, including 2D semiconductors such as MoS\textsubscript{2}~\cite{Sarkar2014} and WS\textsubscript{2}~\cite{Cui2015}.

Ferroelectric FETs (FeFETs) consist of a ferroelectric material such as BaTiO\textsubscript{3} placed between the gate electrode and the source-drain section of the device. MOSFETs are subject to a fundamental limit to the supply voltage, and thus power consumption required to switch at a given temperature, known as the \textit{Boltzmann Tyranny}~\cite{Alam2019}. This related to the steepness of the subthreshold region of a device's IV characteristics and is capped at around 60\.mv/decade at room temperature. A key attraction of these devices is their subthreshold swing which can be below the 60\.mv/decade limit~\cite{Zhang2019}.

\subsubsection{Nonvolatile Memories}
Motivated by the problem of the \textit{von Neumann Bottleneck}, a significant amount of devices research explores the use of novel memory technologies for in-memory computation, in effect, attempting to solve a computer architectures problem using novel devices. However, such novel devices often end up being viewed as less-competitive alternatives to existing (digital CMOS) memory technologies. As we describe in Figure~\ref{figure:DevProcesses}, mapping computational problems to these devices has the potential for a greater research impact.  

The majority of novel memory devices fall under the umbrella of a class of memory called \textit{non-volatile (random access) memory} (NVM), i.e., (random access) memories which retain their state in the absence of a power supply. Table~\ref{table:Memories} shows a comparison of the performance of different novel memory technologies. To this end, we provide a brief overview of novel memory technologies. Chen published a comprehensive review in 2016~\cite{Chen2016}.

\begin{table}[h]
	\caption{A comparison of the performance of emerging NVM technologies. Values are representative and taken from~\cite{Chen2016}.}
	\centering
	\begin{tabular}{c|c|c|c}
		\toprule
		\bf{Memory} & \bf{Switching} & \bf{Cycle} & \bf{Write} \\
		\bf{Technology} & \bf{Speed:} & \bf{Endurance:} & \bf{Voltage} \\
		\midrule
		STTRAM & <\,10\,ns & >\,10\textsuperscript{12} & <\,1\,V \\
		ReRAM & >\,20\,ns & 10\textsuperscript{6}-10\textsuperscript{9} & -0.5-5\,V \\
		PCM & <\,100\,ns & >\,10\textsuperscript{9} & <\,3\,V \\
		FeRAM & >\,20\,ns & 10\textsuperscript{4}-10\textsuperscript{9} & $\approx$\,5\,V \\
		\bottomrule
	\end{tabular}
	\label{table:Memories}
\end{table}

\begin{figure}
	\centering
	\includegraphics[trim= 5.5cm 9cm 5.5cm 10cm, clip=true, width=0.8\linewidth]{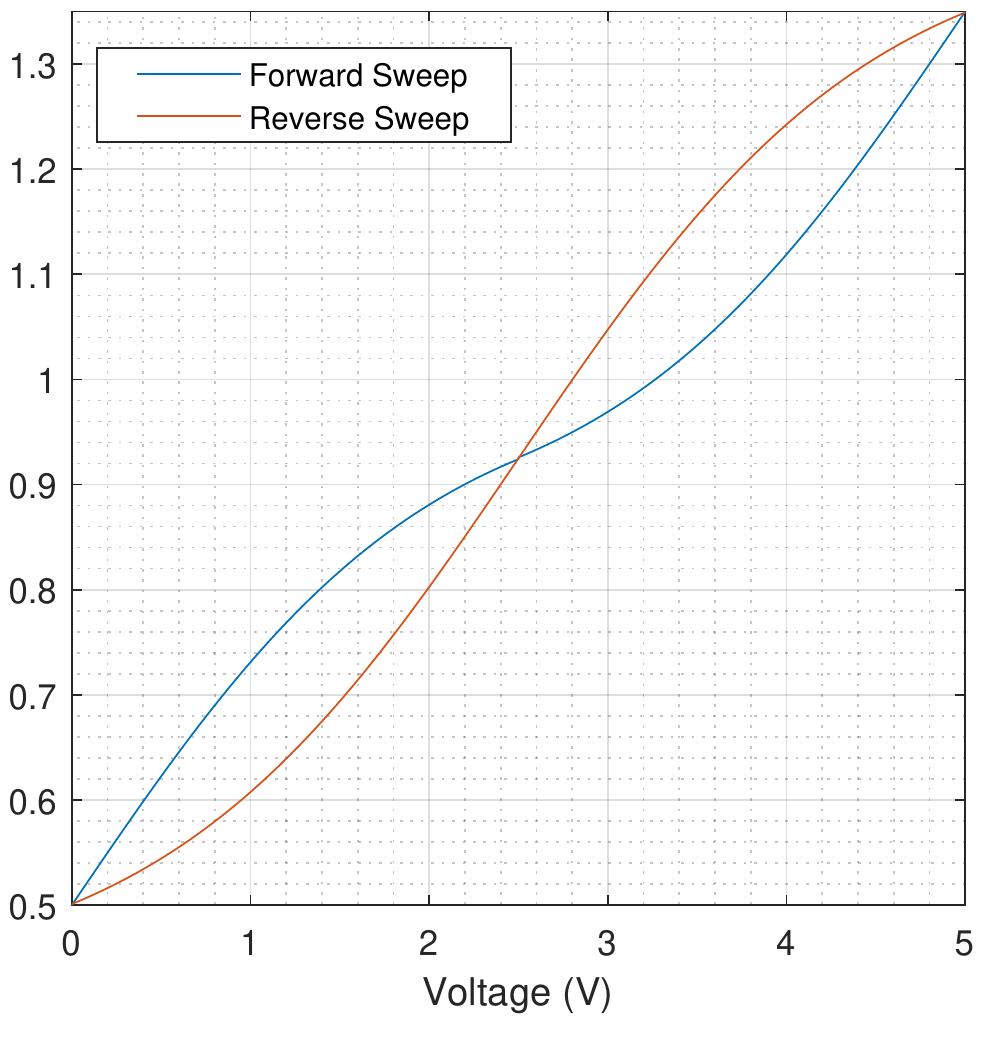}
	\caption{Example of the pinched hysteresis loop which Chua defines as the characteristic property of a memristor~\cite{Chua2014}.}
	\vspace{-0.5cm}
	\label{figure:memristor}
\end{figure}

Among the most widely explored technologies for next-generation computing are ReRAMs. This expression is often used interchangeably with the term \textit{memristor}, however memristors often refer to a specific type of device. First theorised by Leon Chua in 1971~\cite{Chua1971}, the memristor was predicted to be a missing circuit element which relates electrical charge and magnetic flux linkage. Since the first experimental realisation in 2008~\cite{HP2008}, memristors have regularly been touted as a revolutionary device for novel forms of computing and have been the subject of intensive research. Chua updated the definition of a memristor in 2014 to include any device with a pinched hysteresis loop~\cite{Chua2014} (Figure~\ref{figure:memristor}). In this article, we use ReRAM as the general term, and memristor for specific examples.

%Many ReRAM devices fall into two categories: conductive bridge RAM (CBRAM) and valence-change RAM (VCRAM). 
Resistive switching mechanisms in ReRAMs are broadly divided into body-effect and interface-effects~\cite{Guo2020}. Body-effect switching involves a change in the active layer due to different biases, such as the formation of conductive metal filaments (CFs) in the oxide or insulating layer, charge trapping, charge transfer and changes in the carrier concentration. Conductive Bridge RAM (CBRAM) is a common example, with CF formation and rupturing being the source of the high and low resistance states (HRS \& LRS). CBRAM generally requires an initial electroforming process, in which a large voltage is applied to induce a breakdown in the oxide layer that allows for the subsequent formation of CFs. The voltage required to set and reset following this step is typically much lower. However, these CFs are a large source of inter-device variation, as the path and number of filaments that form in a device is difficult, if not impossible, to control. Body-effect ReRAMs also incorporate valence change RAM (VCRAM), where switching is the result of a valence change in the oxide layer, due to oxygen atom migration, leading to conductive path formation and dissipation. Switching in interface-effect ReRAMs is derived from effects at the interface of the active layer and the electrodes, such as Schottky barrier formation, a change in the profile of the potential barrier at the interface or a phase transition at the interface~\cite{Guo2020}. The active layers in such devices typically consist of transition-metal oxides such as HfO\textsubscript{2}~\cite{Tappertzhofen2020}~\cite{Chen2016}. A key advantage of many of these devices is their compatibility with the CMOS process, however non-CMOS devices, such as those based on polymers also exist~\cite{Chen2014}.

%Resistive switching in these devices arises from the formation of oxygen-deficient electrically conductive paths during an initial electroforming process. Conductivity depends on the concentration of oxygen vacancies between the conductive path and high work function electrodes, which modulates a Schottky barrier. %Hybrid devices also exist, such as the plasmonics approach of Di Martino~\cite{DiMartino2016}, or the work of Cho et al.~\cite{Cho2016}, where the resistive states are dependent on Schottky barrier modulation without requiring a forming process.

Phase-change memories (PCMs) are a form of NVM which rely on the phase transition of a given material or materials to change the electronic properties, typically as a result of Joule heating due to an applied current~\cite{LeGallo2020}. Generally, this involves a transition between a crystalline and an amorphous phase, with the former generally representing a low resistance state and the latter a high resistance state~\cite{Chen2016}.

Magnetic or magnetoresistive random-access memories (MRAMs) are a form of NVM in which data is stored in magnetic domains; they use electron spin, rather than charge, to store data. The most common form of this is the magnetic tunnel junction (MTJ), a mature technology which is used most widely in computer hard drive read heads~\cite{Zhu2006}. Traditional MTJs consist of two magnetic layers separated by a non-magnetic or insulating material. The state of the devices is read by measuring their electrical resistance: when the two magnetic layers are magnetised in the same direction, the tunnel current between them is greater and so they are considered to be in a low resistance state. When they are magnetised in opposite directions, the tunnel current is much smaller and so they are considered to be in a high resistance state.

The state of an MTJ is traditionally set by the application of a magnetic field induced by the current in a write line in the device. A more recent form of MRAM is known as spin-transfer torque MRAM (STT-MRAM). This is an MTJ where spin-polarised electrons are used to directly influence the nature of the magnetic domains. If the electrons flowing into a layer change their spin, this induces a torque which is transferred to the nearby layer, reducing the current required to write to the cells.

FeFETs also have utility as NVMs, as the ferroelectric layer retains a permanent polarisation in the absence of a current, meaning the device retains its state. Ferroelectic materials are also used for ferroelectric RAMs (FeRAMs), possesses a similar structure to a DRAM cell, replacing the dielectric layer of the DRAM capacitor with a ferroelectric layer~\cite{Park2018}. These differ slightly from FeFETs in that the transistor need not be ferroelectric.

\subsection{Memory Devices as Solutions to Computational Problems}
\subsubsection{Case Study: Probabilistic Bits}
Though researchers have posited MRAMs as a potential `universal memory'~\cite{Akerman2005}, they also have been proposed for computational applications. One such application is probabilistic bits (P-bits). First proposed by Palem in 2003~\cite{Palem2003}, P-bits are a computational primitive analogous to bits in Boolean logic, or qubits in quantum computing. Unlike bits, which are either 1 or 0 at a given moment, or qubits, which exist in a superposition of states, where they are both 1 and 0, P-bits rapidly flip between 1 and 0 and so have a probability of being 1 or 0 at a given moment, making them a hardware implementation of a Bernoulli random variable. This property allows for them to be used to do computations directly on probabilities.

Khasanvis et al. constructed an array of P-bits based on strained-MTJs (S-MTJs) to implement a Bayesian network~\cite{Khasanvis2015}. Compared to a digital CMOS implementation of the same Bayesian network using a 45\,nm process, Khasanvis et al. claim area reductions up to 127$\times$, a 214$\times$ power reduction and a 70$\times$ lower latency, based on HSPICE simulations comparing S-MTJ based circuits to 5-bit digital CMOS multipliers. However, the precision must also be considered. The 5-bit digital CMOS circuits simulated in the work have a precision of 1/16, whereas the S-MTJ circuits have a precision of 1/10. Component spacing is also an important consideration: Khasanvis et al. calculate a minimum rectangular area of 0.25\,$\mu$m\textsuperscript{2} to avoid magnetic interactions between devices. Modern digital CMOS devices can be below 10\,nm and so have much higher possible integration densities.

Camsari et al. present the example of a \textit{genetic circuit}, used to find the genetic correlation between two siblings (i.e., how related they are). SPICE simulations in which each of the family members are represented by a P-bit device and the correlation is found by using an XNOR gate, along with an RC circuit, to find the time average of the circuit output. Their circuit gave an output of around 0.5, which agrees with Bayes' rule~\cite{Camsari2019}. This is, in effect, a directed graph, which typically requires nodes to be sequentially updated from parent to child, whereas the simulations suggest good asynchronous operation. However, the authors do remark that further work is required to determine whether the behaviour generalises to larger networks.

The key difference between P-bits and MRAMs is that P-bits make use of a typically undesirable property of MTJs: instability. The magnetic materials used in MTJs exhibit a phenomenon that arises in ferro- and ferri-magnetic nanoparticles: superparamagnetism. This is where, under the influence of temperature, the magnetisation of the magnet randomly flips direction. The time period of this flip, $\tau_N$, also known as the Néel relaxation time, is given by the Néel-Arrhenius equation:

\begin{equation}
	\label{equation:Neel}
	\tau_N = \tau_0 \exp^{\frac{E_b}{k_B T}},
\end{equation}
where $E_b$ is the energy barrier level of the magnet, given by the product of the material's magnetic anisotropy energy and its volume. $k_B$ is Boltzmann's constant and $T$ is the temperature. $\tau_0$ is a constant with a value on the order of pico- to nano-seconds. A zero-barrier magnet, where $E_b < k_B T$ would be ideal here, as the time period for a flip would be less than 1\,ns~\cite{Camsari2019}. 

This also illustrates well the intersection of devices, materials and computational research. As P-Bits offer a fundamentally different way to do computations by physically representing probabilities, they map computational problems such as Bayesian networks directly to hardware, which is of interest to both computing and devices researchers. However, as their performance is directly tied to materials properties, the `materials discovery' aspect links directly to the computational problem. 2D materials offer particularly interesting opportunities for P-Bits as they are predicted to have very low magnetic anisotropy energies. The barrier of monolayer CrI\textsubscript{3} has been measured to be as low as 0.66\,meV and is variable with the addition or removal of electrons. A barrier of 0.66\,meV, using Equation~\ref{equation:Neel} corresponds to a $\tau_N$ of about 1\,ns. Another candidate material, strained Fe-doped MoS\textsubscript{2}, has a barrier of about 1.3\,meV~\cite{Chen2015}, corresponding to a $\tau_N$ of 1.06\,ns, which decreases with the addition of strain. For comparison, P-bit approaches using bulk materials from Borders et al. give best-case retention times of a few ms~\cite{Borders2019}, and Mizrahi et al.'s simulations give a natural rate of 518\,Hz, or 1.93\,ms~\cite{Mizrahi2018}. Pileggi et al. report times on the order of a few ns using strained-MTJs~\cite{Bhuin2017}~\cite{Bhuin2017_2}~\cite{Pagliarini2020}, however these require an additional piezoelectric layer, adding material and space costs. 

Camsari et al. do note that magnetic materials are not the only approach to implementing P-bis, proposing a general description of a P-bit as any three-terminal tunable random number generator~\cite{Camsari2019}. The majority of prior implementations of P-bits make use of MTJs, with each P-bit having a structure very similar to cells used for MRAMs, consisting of a transistor with an MTJ cell connected to one of the legs. The input to the device, i.e., the gate, which tunes the device, effectively biases the stochastic output. Johnson-Nyquist noise, as observed in a resistor, is not useful here as it cannot be biased.

\subsubsection{Case Study: Nonvolatile Memories for ANN Inference}
\begin{figure}
	\centering
	\includegraphics[trim= 0cm 0cm 20cm 8cm, clip=true, width=\linewidth]{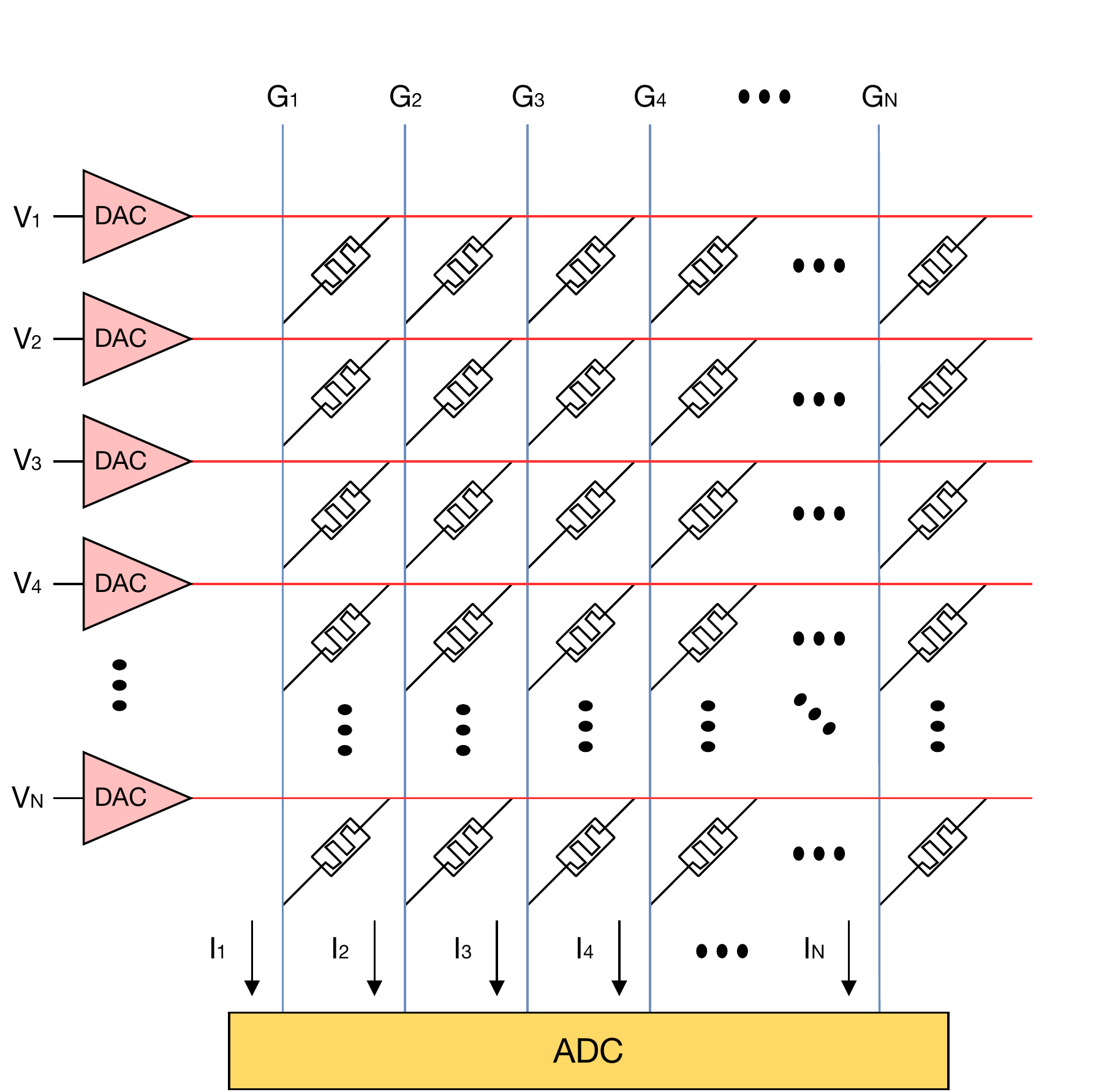}
	\caption{Example of a 1R ReRAM crossbar array. Alternative implementations may have a ReRAM connected to a transistor (1T1R) to allow for individual device selection, or may utilise pairs of ReRAMs in order to encode signed values~\cite{Tsai2018}.}
	\vspace{-0.5cm}
	\label{figure:Crossbar}
\end{figure}

A third example of mapping a computational problem to novel devices is the commonly described use of NVMs in crossbar arrays (Figure~\ref{figure:Crossbar}) to implement MVM operations. These use Ohm's law and Kirchhoff's laws, where the sum of currents from each device gives a final output current. The value of the output current corresponds to a particular signal strength. The major attraction of this approach for computer architects is that it significantly reduces the complexity of the MVM operation. Conventionally, given two vectors, $x$ and $y$, of the same size, $n$, their dot product is $\sum x \cdot y$. If each addition takes one clock cycle, then the complexity of the operation is $\mathcal{O}(n^2)$, i.e., quadratic time. Using the crossbar approach, the entire operation is completed in a single cycle, with complexity $\mathcal{O}(1)$~\cite{Hu2016}, i.e., constant time. 

NVM implementations do face a number of challenges, however. Inherent inter-device variation can become an issue, leading to unpredicatble behaviour. In large arrays, parasitic wire resistance is intensified and restricts the performance of devices. In the case of MVM, this means the precision of the weighted sum is reduced. Li et al. suggest solutions for overcoming issues with inter-device variability and parasitic wire resistance~\cite{Li2018}, stating that variability can be reduced by scaling down devices and wire resistance reduced by 3D integration. Another issue which still persists and is perhaps one of the key hindrances to the scalability of both 2D and 3D memristive circuits is that of sneak path currents~\cite{Seok2014}. Suppression of these requires a selector device with a highly nonlinear current-voltage characteristic connected in series with the ReRAM. This is compounded by the fact that the large nonlinearity results in nonlinear amplification of noise and thus Ohm's law, on which the crossbar MVM operation depends, no longer applies.

NVM-based implementations of CNNs may offer a solution to some of the scalability issues affecting fully connected solutions: they require fewer weights and thus fewer devices, reducing both the influence of parasitic resistances and the precision problems resulting from inter-device variation. Lin et al. recently demonstrated a 3D memristor circuit, consisting of eight monolithically-integrated layers of devices for inference using a CNN~\cite{Lin2020}, in which each device layer represents a layer in the CNN. However, Garbin et al.'s work on memristor-based CNNs demonstrated the requirement for a much higher number of switching cycles, about three orders of magnitude~\cite{Garbin2015} compared to fully-connected approaches, due to the reduced parallelism. Cycle endurance has been a key issue with memristive devices. Yang et al. demonstrated a switching endurance on the order of $10^{10}$ cycles in TaO$_x$-based devices, however many implementations exhibit much lower endurance. 

Modern CPUs operate at switching speeds on the order of GHz, i.e., $10^9$ cycles per second and so even the best of the memristive approaches would not function well here. These circuits are not expected to operate at the same sorts of frequencies, however the endurance is still an important factor. A direct comparison to a CPU is unfair here and so a GPU is worth instead considering, as these are often used for ANN acceleration. A ReRAM-based approach requires one clock cycle per dot-product operation, as does a GPU~\cite{CUDAVMMCycles}. Consider a simple 4x4 dot-product operation. The Nvidia GeForce 210 features 16 CUDA cores operating at 1402\,MHz~\cite{GeForce210}, meaning that roughly 1.4 billion dot-product operations can be completed per second. For a ReRAM implementation of the best-case cycle endurance, operating at a comparable speed, the devices would degrade shortly after one second. Top-end GPUs from Nvidia feature as many as 4608 CUDA cores, as well as \textit{Tensor Cores}~\cite{TitanRTX}, which also offer accelerated performance for ML applications, meaning there is a large disparity between ReRAM and GPU performance. However, parallelised MVM hardware simply transfers the complexity, requiring $\mathcal{O}(n^2)$ devices or cores, instead of $\mathcal{O}(n^2)$ time. As a ReRAM approach requires only one device compared to an entire core in a GPU, ReRAMs offer a significantly more space-efficient solution. 

A final consideration is that not all novel devices are likely to perform equally. For example, certain transistor materials and structures are better-suited to some applications than others; an amplifier circuit design to operate at radio frequencies would perform poorly if one used a transistor designed for audio applications, despite the fundamental circuit structure being the same. Likewise, certain ReRAMs (crossbars) made from a given combination of materials may be more suited to some applications than others. This is another example of how determining one's computational problem first can lead to more impactful research. Flexible or disposable electronics, for example, may not require the endurance of conventional applications, or CMOS processes may be prohibitively expensive or unsuitable for such applications, and so NVMs such as those based on polymers~\cite{Chen2014} would be useful here.

\section{Machine Learning Accelerators}
\label{section:ml-accelerators}

Limitations in the performance of ML using conventional CPUs has resulted in 
the investigation of hardware accelerators, variously referred to by terms including 
`hardware neural networks' or `intelligence processing units'. Often, such approaches are described as \textit{neuromorphic}, 
as their design is influenced in part by biological systems. 
These come in a number of varieties, as the hardware 
requirements depend on the ML approach being implemented. 
It is worth making a distinction here about the precise aspect of 
ML that is being accelerated. By definition, ML involves a learning 
process (training) and is used to make predictions (inference). Thus, accelerators tend to focus either on training or inference.

\subsection{Neuromorphic Systems}
\label{section:NeuroSys}
Neuromorphic computing systems are computing architectures which
draw inspiration from the structure and physical operation of
biological neural architectures. Architectures for neuromorphic
computing range from all-digital implementations~\cite{6055294,
6055293, 6226357, merolla2014million} to mixed analog-digital
implementations which emulate the dynamics of biological neural
systems using the dynamics of man-made structures such as analog
electrical circuits or phase-change materials~\cite{4585796, 6126001,
cruz2013scalable, 0957-4484-24-38-384010,neftci2013synthesizing,
Pfeil1311, wright2013beyond, 6805187}. The models of neural circuits
that neuromorphic computing systems emulate range from simple spiking neural networks to sophisticated models of the ion transport in
biological neurons, captured in millions of differential equation
instances. The applications of these neuromorphic systems demonstrated
in the research literature range from digit recognition, speaker
recognition and sequence prediction~\cite{esser2013cognitive}, to
the modelling of mammalian olfactory
pathways~\cite{10.3389/fnins.2012.00083}.

However, the term `neuromorphic' is not well-defined. For example, a hardware realisation of a DNN using e.g., logic gates to perform MVM might be referred to as a neuromorphic system, despite possessing no real resemblance to a brain. On the other hand, a system with artificial neurons and synapses might also be referred to as neuromorphic. Another issue with discussions of neuromorphic systems is that the current understanding of the brain and animal nervous systems is relatively limited. Thus, in essence researchers are potentially designing systems which do not accurately mimic biological processes, which may lead to unforeseen or faulty behaviour. Much of the modern devices literature relies on the neural network model of McCulloch and Pitts~\cite{McCulloch1943}, and the Hodgkin-Huxley model of spiking neurons~\cite{Hodgkin1952}. Although these have proven successful for neural networks, these are outdated models of neurons, and there are more accurate and up-to-date models, such as the Galves-L{\"o}cherbach stochastic model~\cite{Galves2013}.  Conversely, there may also be unforeseen advantages to such approaches, however there is a large amount of uncertainty, which limits their usefulness in applications beyond research. This, alongside the loose definition of term neuromorphic therefore limits its usefulness in technical discussions, and so we avoid using it.

\subsection{Feedforward ANN Accelerators}

\subsubsection{Digital CMOS}
The reliability and ease of fabrication using CMOS processes makes them 
a natural option for use in ML architectures, as such an approach 
would allow easy and fast integration with existing computer hardware. 
Many CMOS approaches take the form of ASICs or 
FPGAs~\cite{Oh2011}~\cite{Park2013}~\cite{Park2015}~\cite{Gonzalez2017}~\cite{Lee2019}~\cite{Yue2020} 
and a number of commercial options already exist~\cite{IntelMovidius}~\cite{CoralAI}. 

\subsubsection{Nonvolatile Memories}
ReRAM ML accelerators are commonly implemented using a crossbar array, with rows and columns corresponding to a bit or word line. This structure generally acts as the fully-connected layer in an ANN. Operation depends on whether the circuit is being used for training or inference. For training, the resistance of each ReRAM corresponds to the weight of a neuron in the ANN layer and the resistance values are trained either in- or ex-situ. In the former, the best resistance values for each device are determined by adjusting the circuit itself until the error rate is minimised. In the latter case, the optimum resistance values are determined outside of the circuit, e.g., through simulation, and the final values are programmed in before the circuit is used.

ReRAM crossbars are typically implemented as a single component connected to external off-the-rack circuitry, such as selectors, sense amplifiers, digital-to-analogue (DAC) and analogue-to-digital (ADC) converters. This external circuitry can become a limiting factor in the gains made by the ReRAM approach. If the external circuitry requires a significant amount of real estate or power consumption, then the performance gains of ReRAM crossbars may be significantly attenuated. 

The inherent inter-device variation can also become an issue in ReRAM approaches. In large arrays, parasitic wire resistance is intensified and restricts the performance of devices. In the case of a MVM, this means the precision of the weighted sum is reduced. Li et al. suggest solutions for overcoming issues with inter-device variability and parasitic wire resistance~\cite{Li2018}, stating that variability can be reduced by scaling down devices and wire resistance reduced by 3D integration. Another issue which still persists and is perhaps one of the key hindrances to the scalability of both 2D and 3D ReRAM circuits is that of sneak path currents~\cite{Seok2014}. Suppression of these requires a selector device with a highly nonlinear current-voltage characteristic connected in series with the ReRAM. This is compounded by the fact that the large nonlinearity results in nonlinear amplification of noise and thus Ohm's law, on which the crossbar MVM operation depends, no longer applies.

\subsubsection{Other Feedforward ANN Technologies}
Though CMOS and ReRAM implementations are the most common approach to implementing feedforward ANNs, they are not the only approach. \textit{Neurotransistors} have also been explored. These utilise a ReRAM integrated with a transistor gate, which mimics the `integrate-and-fire' behaviour of biological neurons~\cite{Wang2018}~\cite{Baek2020}.

Optical implementations of matrix multiplication were demonstrated at least as early as 1970~\cite{Heinz1970} and a large body of the work on optical computing focuses on parallel computing and particularly matrix multiplication, making it of obvious interest for ML applications. However, optical/photonic implementations of neural networks specifically also exist. Hamerly et al. demonstrated an all-optical neural network which offered very low-energy operation for MVM operations and gigahertz speeds, as well as $\leq 1$\% error rates on MNIST~\cite{Hamerly2019}.

We would also be remiss not to discuss \textit{Reservoir Computing}, a computational framework based on recurrent neural networks (RNNs)~\cite{Tanaka2019}. RNNs are a type of ANN which have a temporal sequence, and are used for tasks such as natural language processing (NLP)~\cite{Li2015}. RNNs use previous outputs as inputs, i.e., the activation at a given time is a function of the activation at a previous time.

A reservoir computer (RC) builds upon several RNN models and so is also suited for temporal data processing. The reservoir part of a RC is treated as a black box, but has two requirements: it must be made up of individual nonlinear units and be able to store information~\cite{Soriano2017}. A RC uses the reservoir to map inputs into a higher dimensional computing space, and then conducts pattern analysis in a readout section. Unlike other ANN approaches, a reservoir computer does not train the weights of the input or reservoir sections, only the readout part. This theoretically means a simplified and faster training process, as simple training algorithms, such as linear regression, can be used, with a focus on reduced computational cost compared to alternate approaches~\cite{Tanaka2019}. As with accelerators for other ML approaches, RCs have been realised using different devices and materials, including a photonics-based approach~\cite{Larger2017}, or even a literal reservoir in the form of a water tank~\cite{Fernando2003}. For a comprehensive review, see Tanaka et al.~\cite{Tanaka2019}.

\subsection{Spiking Neural Network Accelerators}
\subsubsection{Artificial Synapses}
\begin{figure}
	\centering
	\includegraphics[trim = 0cm 18cm 8cm 0cm,clip=true, width=0.7\linewidth]{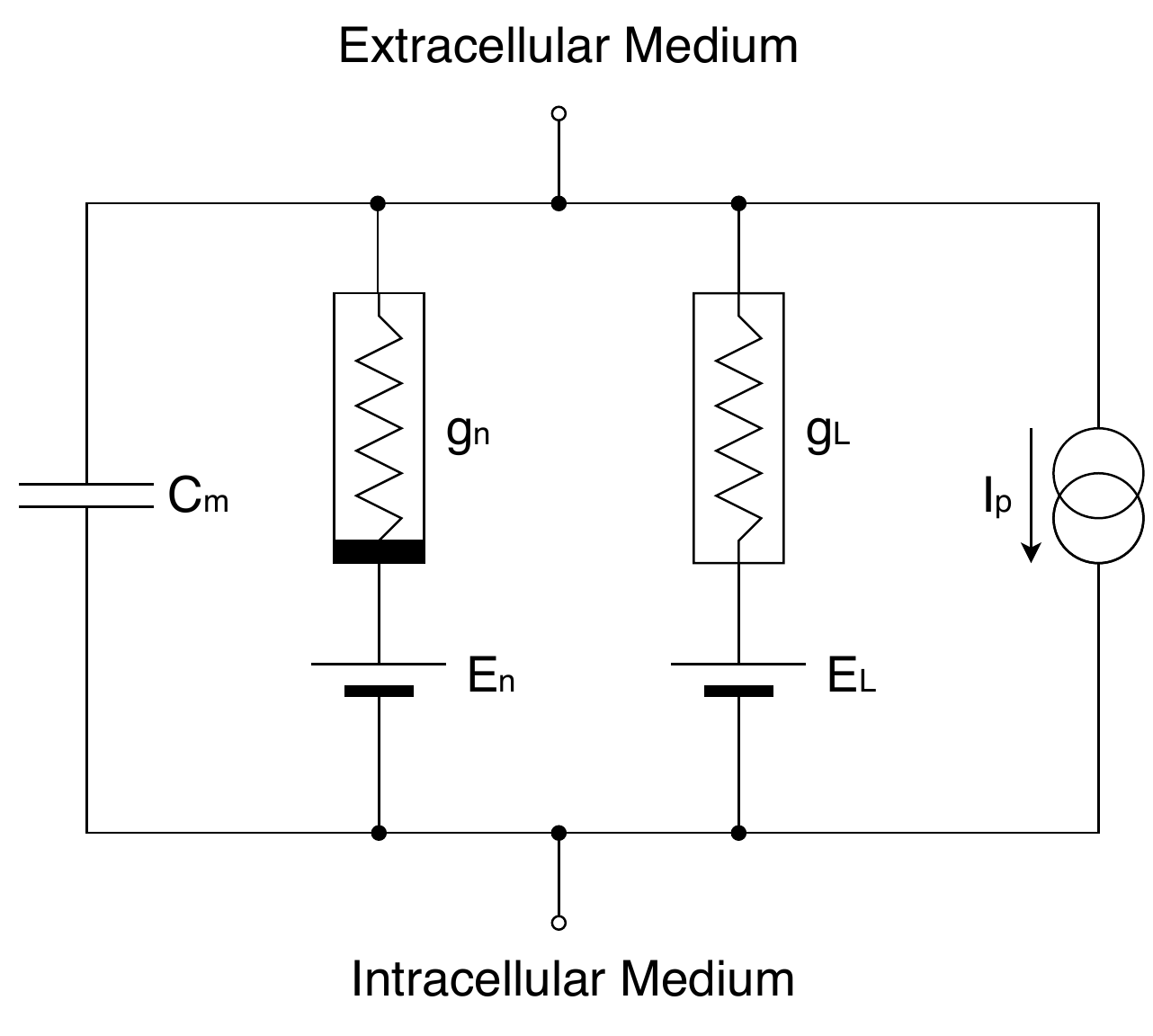}
	\caption{Schematic of the Hodgkin-Huxley model of a neuron~\cite{Hodgkin1952}. $C_m$ is a capacitance representing a lipid bilayer, $g_n$ is a nonlinear voltage source representing a voltage-gated ion channel (dependent on both voltage and time), $g_L$ is a linear voltage source representing the leak channels, $I_P$ is a current source representing ion pumps and the electrochemical gradients are represented by the voltage sources $E_n$ and $E_L$.}
	\vspace{-0.5cm}
	\label{figure:ArtSyn}
\end{figure}

As discussed in Section~\ref{section:NNs}, SNNs more closely resemble biological systems than feedforward ANNs. The spike train discussed serves as the integrate-and-fire mechanism, whereby the inputs to a neuron are summed until some threshold is reached, after which they fire. A common model, the Hodgkin-Huxley model, describes the activation level of a neuron with sets of differential equations with tens of parameters, which are difficult to compute~\cite{Hodgkin1952}. Figure~\ref{figure:ArtSyn} shows a schematic of the model.

`Leaky' models are preferable, as these are less difficult to compute. A simple example of such a leaky neuron would be a leaky capacitor which sums the currents from input synapses and whose leakage current brings the neuron to a rest state. Alternatives using existing devices, such as Schmitt triggers~\cite{Kim2019_2}, also exist. 

Artificial synapses are themselves an area of active research, with many articles published demonstrating synapse-like performance for a range of devices and materials. A comprehensive review of nanoelectronic materials for use as artificial synapses was recently published by Sangwan and Hersham~\cite{Sangwan2020}. 

\subsubsection{ReRAM as Artifical Synapses vs ReRAM as Feedforward Accelerators}
ReRAM are often discussed in two main applications for ML approaches: as artificial synapses and as computational devices for accelerating feedforward ANNs. This can potentially lead to confusion in terminology. Often people may refer to the nodes in an ANN as neurons. As this borrows from neuroscience terminology, it can be easy to conflate and associate research into ANN accelerators as being aligned with artificial synapses, where this is not always the case. Section~\ref{section:computing-devices} presented the basic operation of a ReRAM: the resistance is dependent on the current that has previously flowed through it, an integral with respect to time. This can be seen as analogous to the operation of a biological neuron, in which charges in the form of ions accumulate over time until the activation potential (i.e., threshold) is reached. When this threshold is reached, the neuron fires. This uses the properties of ReRAM in a different way to a crossbar MVM accelerator, where the state of the ReRAM represents a stored weight or value.

\subsubsection{Digital CMOS SNN Accelerators}
The is also a large body of research into hardware implementations of SNNs. Perhaps the best-publicised of these implementations is IBM's \textit{TrueNorth}, a hardware SNN implemented using 5.4 billion transistors and fabricated on a Samsung 28\,nm CMOS process~\cite{TrueNorth}. TrueNorth has a claimed power consumption of about 65\,mW for multi-object detection and classification with 240$\times$400 pixel 30 frames-per-second video input. However, their claimed classification precision of 0.85 on the test set of the DARPA Neovision2 Tower dataset was for 1920$\times$1088 pixel images, making comparisons between the reported power consumption and the associated accuracy difficult to establish. In terms of power density, TrueNorth also reports a power density of 20\,mW cm\textsuperscript{-2}, compared to 50-100\,W cm\textsuperscript{-2} for CPUs. For example, SpiNNaker (Spiking Neural Network Architecture), a massively parallel manycore architecture and part of the \textit{Human Brain Project}, currently consists of 57,600 ARM9 18-core processors, corresponding to a total of 1,036,800 cores~\cite{Spinnaker:2020}. The TrueNorth team compare their platform to SpiNNaker, reporting 769$\times$ less energy consumption and 11.4$\times$ less silicon area, however these comparisons are not like-for-like: SpiNNaker cores are fabricated on a 130\,nm process~\cite{SpinnakerChip} and different benchmarking suites are used.

It is important to draw a distinction between systems such as SpiNNaker, TrueNorth and ASIC approaches such as Google's TPUs~\cite{Google:TPU}. Although all are based on CMOS processes, their constructions and purposes are different. SpiNNaker aims to simulate the human brain, whereas TrueNorth has dedicated CMOS neurons and synapses, which do computation inspired by the way the brain works. ASIC approaches such as TPUs are simply hardware implementations of ANNs, e.g., to accelerate matrix multiplication.

\subsection{Acceleration of other ML Approaches}
As discussed in Section~\ref{section:ml}, there are a plethora of ML approaches beyond ANNs. Many of these are less-explored than ANNs however, so the body of work on hardware acceleration of them is also correspondingly smaller.

\subsubsection{Support Vector Machines}
Many hardware implementations of SVMs appear to be based on FPGAs~\cite{Irick2008}~\cite{Kyrkou2012}~\cite{Jallad2014}~\cite{Zhao2017}~\cite{Wang2018SVM}, though most non-FPGA approaches still utilise CMOS technology. Genov and Cauwenberghs demonstrated a silicon-based `kerneltron' SVM processor, which is a mixed-signal system consisting of an array of charge-injection devices combined with DRAM~\cite{Genov2003}. Kang and Shibata demonstrated a Gaussian-kernel based SVM using 180\,nm CMOS technology~\cite{Kang2010} with a reported power consumption of around 220\,$\mu$W. Takagi et al. reported a sub-100\,mW SVM accelerator using a 65\,nm CMOS process to implement a Histogram of Oriented Gradients algorithm~\cite{Takagi2013} and Jeon et al. fabricated an accelerator for facial recognition using 5-transistor memory cells fabricated on a 40\,nm CMOS process, with a reported power consumption of 23\,mW~\cite{Jeon2015}.

\subsubsection{Decision Trees}
Hardware implementations of DTs are suggested to offer significant savings in power and improvements in throughput; de Franca et al. demonstrated a hardware DT classifier for dealing with network attacks with a reported 15$\times$ throughput compared to a software approach written in C++ which also had an energy consumption of 0.03\% of the corresponding software implementation~\cite{deFranca2014}. As with SVMs and indeed most ML accelerators, initial work on DT hardware accelerators utilised FPGAs. Struharik proposed architectures suitable for implementation in both FPGAs and ASICs~\cite{Struharik2011}. The hardware approaches proposed offered speedups between 8.58$\times$ and 1790.4$\times$ that of comparable software approaches whilst also offering reductions in the hardware resources required.

Saqib et al. proposed a pipelined accelerator for DT inference, testing it on the Iris and Contact Lenses databases and achieving accuracies of 98\% and 83.3\% respectively~\cite{Saqib2015}.

In-memory approaches have also attracted interest. Kim et al. proposed a DT accelerator design using memristors~\cite{Kim2019} for image-recognition applications. Their design was simulated for a 45\,nm CMOS process and evaluated using several benchmarks, including MNIST, where it achieved an accuracy of 97.5\%. 

Kang et al. used a bespoke 6-transistor SRAM-based IC fabricated on a 65\,nm CMOS process to implement a Random Forest DT accelerator~\cite{Kang2018}.

More recently, Goswami et al. demonstrated a `molecular memristor' for hardware implementation of a decision tree~\cite{Goswami2021}. As well as demonstrating ML acceleration using a novel device, this work also uses materials properties to implement logic functions, with the different redox states of a given device representing different logic functions.

\subsection{Hardware vs Software}
It is worthwhile now to return to the thesis of this article, the disconnect between the materials devices communities on one hand, and the computer science and computer hardware architectures communities on the other. As discussed in Section~\ref{section:ml}, a number of benchmarks exist for evaluating the performance of ML systems. A comparison between the performance of new circuits implemented using newly-proposed materials and devices versus circuits based on silicon (often CMOS) and digital logic is a good way to address this gap, as metrics of success vary between the disciplines.

\begin{figure}[h]
	\centering
	\includegraphics[trim= 4.75cm 9.5cm 5cm 10cm, clip=true, width=\linewidth]{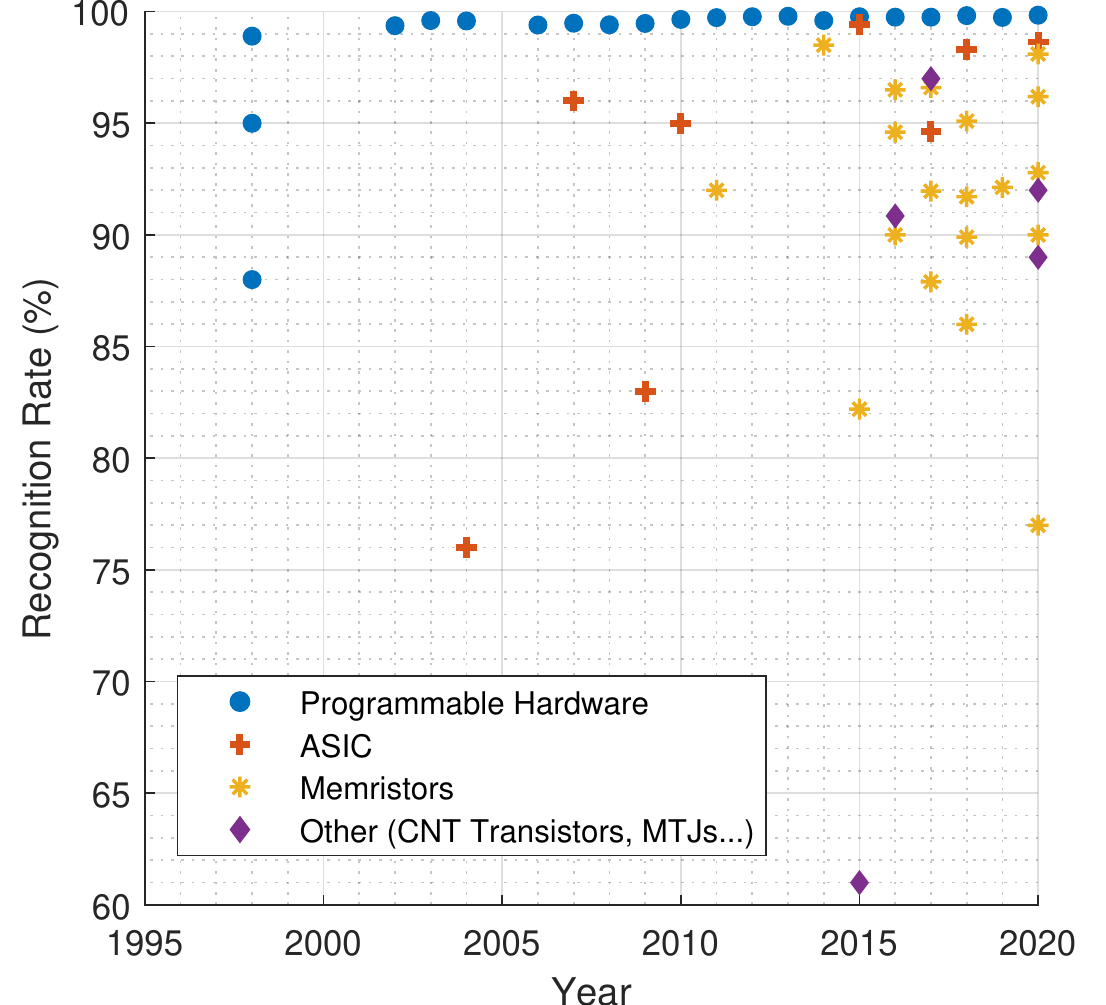}
	\vspace{-0.5cm}
	\caption{Scatter plot showing the reported accuracies on the MNIST dataset of several technologies currently being researched.}
	\vspace{-0.2cm}
	\label{figure:MNIST}
\end{figure}

As we have already noted, neural networks implemented using novel materials and devices often perform significantly worse than their digital and silicon CMOS counterparts. Figure~\ref{figure:MNIST} shows the accuracies of several digital and silicon CMOS-based and approaches based on novel materials and devices against a given year. Tables~\ref{table:MNISTHardware} and~\ref{table:MNISTSoftware} show the corresponding specifications of each system. Again, these are intended as representative rather than exhaustive comparisons.

\begin{table*}[t]
	\caption{Performance of selected ReRAM-based neural networks, the year of the work
			 and their reported MNIST accuracy.}
	\centering
	\begin{tabular}{c|c|c|c|c}
		\toprule
		\bf{ANN Type}	& \bf{ReRAM Material} & \bf{Year} & \bf{MNIST Recognition Rate}	& \bf{Ref.} \\
		\midrule
%		- &  &	& 	&  \\
%		SNN (Simulation) & - & 2011	& 92.00\%	& \cite{Querlioz2011} \\
		DNN & OxRAM & 2014	& 98.50\%	& \cite{Garbin2014} \\
		CNN & PCM & 2015	& 82.20\%	& \cite{Burr2015} \\
		CNN & AlO\textsubscript{x}/HfO\textsubscript{2} & 2016	& 90.00\%	& \cite{Woo2016}  	\\
		Binary NN & TaO\textsubscript{x}/HfO\textsubscript{2} RRAM & 2016	& 96.50\%	& \cite{Yu2016} \\
%		- & - & 2016	& 94.60\%	& \cite{Liu2016}	\\
%		DNN (Simulation) & - & 2017	& 91.96\%	& \cite{Hasan2017}	\\
%		Feedforward (Simulation) & - & 2017	& 96.60\%	& \cite{Liu2017} 	\\
		1-Layer NN & SiN Memristor & 2017	& 87.90\%	& \cite{Kim2017}	\\
		1-Layer NN & 1T1R TaO\textsubscript{x}/HfO\textsubscript{2}/Pd RRAM & 2018	& 89.90\%	& \cite{Hu2018} 	\\
		MLP & SiGe epiRAM & 2018	& 95.10\%	& \cite{Choi2018}	\\
		2-Layer NN & Ta\textsubscript{x}/HfO\textsubscript{2}/M (M=Pt, Pd) & 2018	& 91.71\%	& \cite{Li2018_Conf}		\\
		3-Layer NN & 130\,nm CMOS & 2018	& 86.00\%	& \cite{Jiang2018}	\\
		CNN & 1T1R Pt/TaO\textsubscript{x}/Ta & 2019	& 92.13\%	& \cite{Wang2019}	\\
		CNN & Pt/HfO\textsubscript{2}/Ta MRAM & 2020	& 98.10\%	& \cite{Lin2020}	\\
		5-Layer CNN & 1T1R TiN/TaO\textsubscript{x}/HfO\textsubscript{x}/TiN & 2020	& 96.19\%	& \cite{Yao2020}	\\
		CNN & 1T1R Ti/HfO\textsubscript{2}/TiN MRAM & 2020	& 92.79\%	& \cite{Chen2020}	\\
		SNN & Ta/Ta\textsubscript{2}O\textsubscript{5}/AlN/Graphene & 2020	& 77.00\%	& \cite{Yan2020}	\\
		1-Layer NN & Amorphous InGaZnO 1T2R & 2020	& 90.00\%	& \cite{Jang2020}	\\
		\bottomrule
	\end{tabular}
	\label{table:MNISTHardware}
\end{table*}

\begin{table}
	\caption{Performance of selected  neural networks based on programmable digital hardware (e.g., CPU or GPU), the year of the work
		and their reported MNIST accuracy.}
	\centering
	\begin{tabular}{c|c|c|c}
		\toprule
		\bf{Algorithm/}	&  & \bf{MNIST}	&  \\
		\textbf{Model} & \bf{Year} & \bf{Recognition} & \bf{Ref.} \\
		 &  & \bf{Rate} & \\
		\midrule
		Linear Classifier 1-Layer NN  & 1998	& 88.00\%	&   \cite{Lecun1998}	\\
		$k$-NN, Euclidean (L2) & 1998	& 95.00\%	& 	\cite{Lecun1998} \\
		LeNet-4 CNN & 1998 & 98.90\%	& 	\cite{Lecun1998} \\
		3-NN &	2002 	& 99.63\%	& \cite{Belongie2002} \\		
		2-layer NN  & 2003	& 99.60\%	& 	\cite{Simard2003}	\\
		Limited Receptive Area  &	2004 	& 99.58\%	& \cite{Kussul2004} \\
		CNN & 2006	& 99.40\%	& 	\cite{Ranzato2006} 	\\
		Nonlinear Image Deformation &	2007 	& 99.48\%	& \cite{Keysers2007} \\
		SVM &	2008 	& 99.41\%	& \cite{Labusch2008} \\
		CNN &	2009 	& 99.47\%	& \cite{Jarrett2009} \\
		Deep-CNN &	2010 	& 99.65\%	& \cite{Ciresan2010} \\
		Deep-CNN &	2011 	& 99.75\%	& \cite{Ciresan2011} \\
		Multi-Column DNN &	2012 	& 99.77\%	& \cite{Ciresan2012} \\
		DropConnect & 2013	& 99.79\%	& 	\cite{Wan2013}	\\
		Deeply-Supervised Nets & 2014	& 99.60\%	&  	\cite{Lee2014}	\\
		APAC &	2015 	& 99.77\%	& \cite{Sato2015} \\
		13-layer ANN &	2016 	& 99.75\%	& \cite{Hasanpour2016} \\
		Capsule Vectors &	2017 	& 99.75\%	& \cite{Sabour2017} \\
		Random Multimodel Deep Learning & 2018	& 99.82\%	& 	\cite{Kowsari2018}	\\
		Local Error Signals &	2019 	& 99.74\%	& \cite{Nokland2019} \\
		Branching/Merging CNN + & 2020	& 99.84\%	& 	\cite{Byerly2020}	\\
		Homogeneous Filter Capsules  & & & \\
		\bottomrule
	\end{tabular}
	\label{table:MNISTSoftware}
\end{table}

The low accuracy of approaches based on novel materials and devices when compared to approaches based on digital silicon- and CMOS-based hardware, including software-programmable processors and GPUs, or fixed-function ASICs and other accelerators, is immediately obvious. The accuracies of ASIC approaches seem to lag algorithmic approaches by about ten years, despite ASIC being a mature technology. The ASIC and programmable hardware approaches both show a trend of progress, i.e., they become more accurate over time, however this does not seem to be as true for approaches based on ReRAMs or other experimental materials and devices, where time does not seem to correlate with improved accuracy. As the `other' approaches are not a unified technology, but rather a range of different implementations, this can be forgiven, as a comparison would not make sense. These are included instead to indicate the accuracies reported for approaches outside the most popular ones. For memristors, a ten-year lag might be understandable, given the relative youth of the technology compared to CMOS or software (memristors were first experimentally realised in 2008~\cite{HP2008}, ten years after the MNIST dataset was released~\cite{Lecun1998}), however there is a lack of a clear trend in increasing accuracy over time.

Several authors have remarked on the scarcity of full-scale physical realisations of complete systems, and expressed the importance of experimental demonstrations of these~\cite{Hu2018}~\cite{Li2018_Conf}. We suggest a parallel research path at each level of abstraction will have significant impact. Further materials research will help identify the best materials combinations for different applications. For example, 2D materials exhibit resistive switching mechanisms that do not exist in conventional bulk materials. Engineering of these properties may allow for greater control of switching, and thus more reliable and durable devices. Li et al.~\cite{Li2021} recently demonstrated GRM-based ReRAMs using layers of PdSe\textsubscript{2}, where resistive switching emerges due to grain boundary formation resulting from local phase transitions induced by electron beam irradiation. This control of grain boundaries allows for guided filament formation, allowing greater control than the stochastic filament formation typically observed in conventional materials. Alongside this, progress towards full-scale physical realisations will aid in overcoming the engineering challenges that frequently emerge with producing novel devices at scale, as well as offering more tangible evidence for the impact and advantages of a given approach. Lastly, suitable application of novel devices can allow for more immediate application of novel technologies. For example, some approaches may eliminate the need for large numbers or high-endurance devices, as we discuss in Section~\ref{section:computing-devices}.

The perceptions of accuracy in each community are also worth highlighting. It is not uncommon in devices research articles to see claims that accuracies of 90\,\% are comparable to a simulated hardware approach accuracy of 94\,\%, for example. Initially, this might appear relatively close, however, when the difference between best and the tenth best digital approaches have a difference of only 0.09\,\%, a 4.4\,\% difference is almost 50$\times$ larger. Of course, the raw numbers do not tell the entire story. The leading digital approaches have millions of parameters; the top algorithm at the time of writing has 1,514,187 trainable parameters~\cite{Byerly2020}. In comparison, many implementations based on novel materials and devices have only hundreds or thousands and so, understandably, do not perform as well. This makes it clear that a better metric is required, which we discuss in Section~\ref{section:outlook}. 

\subsection{Materials, Devices, Architectures, and Algorithms}
The overview of computing devices, algorithms, complexity and architectures we have presented, alongside comparisons of ML algorithm implementations in digital silicon- and CMOS-based hardware, as well as those based on novel materials and devices, brings us to a key thesis of this work. Many research articles published in the devices literature refer to phenomena such as the `decline' of CMOS and indeed appear to treat CMOS circuits as synonymous with digital circuits. This is not the case, however. Many novel devices, including ReRAMs, are fabricated using CMOS processes, and some of the best-performing hardware implementations of ML algorithms, such as IBM's \textit{TrueNorth}~\cite{TrueNorth}, use CMOS processes. It is clear then that CMOS technology is not the limiting factor here, but the architecture. Architectures are thus equally as important as the fundamental technology, be it CMOS or some alternative. Despite being published in 1985, Keyes' observations~\cite{Keyes1985} about the successes of CMOS remain true and CMOS remains the most successful technology for computing devices, regardless of the high-level computer architecture. Simply recreating CMOS devices using novel materials, whilst maintaining the same fundamental way of doing computation has limited impact, as few match or better CMOS in the criteria outlined by Keyes. It is too early to say whether, under alternative computing paradigms, CMOS technology will remain dominant and so it remains worthwhile to investigate alternative technologies. A hybrid approach, where new materials and devices are integrated into CMOS processes is one possible avenue. Alternatively, technologies such as P-bits, as discussed in Section~\ref{section:computing-devices}, where the fundamental nature of the computation relies on completely different device and material properties, mean it is likely that CMOS will not be the best technology for all applications.
\section{Outlook for Devices Research}
\label{section:outlook}

\subsection{Computational Complexity of Neural Networks}
Livni et al. addressed the complexity of training neural networks~\cite{Livni2014}. The authors remark that successful learning with neural networks is computationally hard (i.e., is difficult to solve efficiently in polynomial time) in the worst-case, however, several tricks exist which allow for improvement in the efficiency of successful training.

The first and perhaps most relevant of these tricks to device physicists is to consider the activation function. Activation functions are a key part of ANNs, determining the threshold at which a neuron fires. Table~\ref{table:activations} shows the most commonly used. Here we can see that different functions have different complexities and thus will have an influence on the overall efficiency of the network. If the wrong activation function is chosen for a given ANN hardware implementation, then it may severely limit the performance gains. However, a hardware implementation of these may offer significant advantages: if the transfer characteristic of a given device operates in a single clock cycle, then any of the activation functions will be reduced to a complexity of $\mathcal{O}(1)$.

\begin{table}[h]
	\caption{Common activation functions for ANNs and their associated (worst-case) complexity. $M(n)$ represents
	the complexity of a given multiplication algorithm.}
	\vspace{-0.1in}
	\centering
	\begin{tabular}{c|c|c}
		\toprule
		\bf{Function}	 & \bf{Equation}	& \bf{Complexity} \\
		\midrule
		Identity			   & $f(x) = x$		& $\mathcal{O}(n)$	\\
		Binary Step 			& $f(x) = \begin{cases} 0,  & \text{for}\ x > 0 \\ 1, & \text{for}\ x \geq 0 \end{cases}$		& $\mathcal{O}(1)$		 \\
		Tanh				 & $f(x) = \tanh(x)$		& $\mathcal{O}(M(n)log(n))$\textsuperscript{1}	\\
		ReLU				& $f(x) = \begin{cases} 0,  & \text{for}\ x \leq 0 \\ x, & \text{for}\ x> 0 \end{cases}$		& $\mathcal{O}(n)$	\\
		Sigmoid			 & $f(x) = \frac{1}{1 + \exp^{-x}}$		& $\mathcal{O}(M(n)log(n))$\textsuperscript{1}	\\
		ArcTan 			& $f(x) = \arctan(x)$		& $\mathcal{O}(M(n)log(n))$\textsuperscript{1}	 \\
		\bottomrule
	\end{tabular}
	\\ \vspace{0.2cm}
	\textsuperscript{1} Using Arithmetic–geometric mean iteration~\cite{Brent1976}.
	\label{table:activations}
\end{table}

Some work exists in this domain already: Oh et al. recently demonstrated a memristor crossbar which incorporated a VaO\textsubscript{2} device as a ReLU activation neuron~\cite{Oh2021}, however there does not yet seem to be a significant focus in this area, despite the potentially large impact. For example, Mennel et al. demonstrated machine vision using GRM-based photodiodes in which an image sensor functions simultaneously as a ANN~\cite{Mennel2020}, offering promise for reduced hardware footprints. However, they implement the activation function off-chip, which introduces communication and hardware overheads. A fully-integrated solution would further improve the impact of such an approach.

\subsection{Figures of Merit}
Establishing standard figures of merit to evaluate and compare different approaches is vital. Raw power/energy consumption alone is not a suitable metric, as highlighted in the discussion of computational complexity, as a faster/more efficient device may be used in a less efficient architecture. We have also seen that, for standard benchmarks such as MNIST, simply reporting the classification accuracy is not a good measure, as the best-performing digital approaches utilise millions of parameters and thus both billions of transistors and billions of clock cycles. When compared to an experimental device fabricated from novel materials and devices with perhaps a few hundred or a few thousand devices, the comparison becomes unfair. Thus we suggest that it is better to evaluate systems as black boxes, where the relations between outputs are used as performance indicators. We consider a general accelerator as a black box with the following output information available:

\begin{itemize}
	\item Device Count: the total number of devices (e.g., transistors, ReRAMs, or some yet-to-be-invented device);
	\item Power Draw: the average power consumed by the system;
	\item Recognition Rate: the prediction accuracy of the system (i.e., the number of correct predictions divided by the total number of examples);
	\item Runtime: the time taken to make a prediction. 
\end{itemize}

Based on these, we propose the following figures of merit for ML accelerator approaches.

\subsubsection{Joules Per Recognition}
As discussed already, many of the best-performing digital logic approaches for MNIST utilise over a million different parameters and many layers. This corresponds to a high cost in both time, resources, and energy consumption. By comparison, approaches based on novel materials and devices, such as ReRAM crossbars, use significantly fewer resources, however have lower recognition rates. Ideally, the higher the recognition rate the better, but the time and power costs can offset the advantages of accuracy: a facial recognition algorithm that only unlocks a mobile phone only when it is 99.98\% certain the right face has been scanned is not much good if the algorithm takes several minutes and depletes the battery. On the other hand, an ML algorithm used for medical diagnosis must be as accurate as possible in order to minimise false positives or false negatives, although if it takes several weeks to run, this may also be a problem. \textit{Joules per Recognition} (JPR) is thus a good performance metric for a system, as it accounts for energy consumption, time and accuracy:

\begin{equation}
	\textrm{JPR} = \frac{\textrm{Power}\times\textrm{Time}}{\textrm{Recognition Rate}}.
\end{equation}

\subsubsection{Devices Per Recognition}
Energy consumption alone does not tell the whole picture however: if one requires significantly more devices to run an algorithm with equivalent accuracy and in the same time, but with reduced power consumption, then the runtime power savings may be offset or negated by the physical space required by the system and, more generally, by manufacturing costs and energy requirements. Thus, \textit{Devices per Recognition} (DPR) is a useful metric, as it evaluates the accuracy of the system against the number or physical devices required to implement it:

\begin{equation}
	\textrm{DPR} = \frac{\textrm{Device Count}}{\textrm{Recognition Rate}}.
\end{equation}

This metric is also useful as it helps quantify the device needed for a given task; a given novel device will not be a good candidate for computing tasks if the number of devices needed to solve realistic problem sizes is significantly large. Likewise, if a solution to a computational problem using some novel device requires only tens of devices, with comparable performance to a computer program running on reprogrammable digital CMOS hardware, using millions of transistors, then the novel solution would be a clear leader in this case. 

\subsubsection{Dot-Product Operations Per Second}
The final figure of merit we propose is \textit{dot-product operations per second} (DPOPS):

\begin{equation}
	\textrm{DPOPS} = \frac{\textrm{Dot-Product Operations}}{\textrm{Time}}.
\end{equation}

This is analogous to floating point operations per second, a common metric for computer performance. DPOPS describes the throughput of an accelerator in terms of speed and MVM size. Each crosspoint in a given circuit corresponds to a single weight, and so the number of operations in a single cycle would be equivalent to the number of crosspoints. The limiting factors here are thus crossbar size and also the time taken for a crossbar read and write operation; devices which are slow to program will limit the sequential speed of the system.

\subsubsection{Optimisation of System Parameters}
The desirable or acceptable performance of a system is ultimately application-specific and so it is likely that a balancing of various parameters will be necessary for each individual system design. Multi-objective or Pareto optimisation with our proposed figures of merit would provide a useful framework with which to determine the ideal balance of different parameters in a given implementation. 

\subsection{Other Proposals to Facilitate Evaluation and Comparison Between Domains}
For reasons we discuss in Section~\ref{section:complexity}, fixed-function hardware such as Google's TPU have begun to use the IEEE \texttt{bfloat16} format, and so we propose that this be a common standard of precision used for ML systems.

A final suggestion we propose is the standardised reporting of the computational cost in devices papers, e.g., as a dedicated section. Not only will this help researchers consider the scalability and viability of their own research, but it will help readers and those who build upon their work to better direct subsequent investigations. In the textbook, \textit{Introduction to Algorithms}, the authors discuss `algorithms as a technology'~\cite{Cormen2009} and this is an important consideration. Returning to the example of ReRAM-based MVM operations, we can see that changing the architecture reduces the time complexity of the operation, but, more importantly, this does not smuggle the complexity elsewhere: the resource complexity grows linearly with input size.

For ML approaches, whether based on digital logic implemented in CMOS, or based on digital- or analog-domain computation using novel materials and devices, the understanding that any ML algorithm has three components makes this easy to calculate, as the complexity of the system or algorithm will be whichever term in the \textit{Learning = Representation + Evaluation + Optimisation} equation grows the fastest.

\subsection{Machine Learning Computational Problems and Hardware Solutions}
In light of the end of Dennard scaling around 2006, researchers at UC Berkeley~\cite{BerkeleyDwarfs} from a variety of fields met to discuss and make predictions for the transition towards parallel computing. They described thirteen `dwarfs', which are algorithmic methods that describe patterns of data communication and computation. The classification also accounts for ML applications, and Table~\ref{table:MLDwarfs} lists the thirteen dwarfs and their corresponding applications in ML. 

\begin{table}[h]
	\caption{The thirteen computational dwarfs identified by UC Berkeley Researchers~\cite{BerkeleyDwarfs} and their corresponding applications in ML. Abbreviations: PCA (principal component analysis); ICA (independent component analysis).}
	\centering
	\begin{tabular}{c|c}
		\toprule
		\bf{Dwarf} & \bf{ML Application} \\
		\midrule
		Dense Linear Algebra & SVMs, PCA, ICA \\
		Sparse Linear Algebra & SVMs, PCA, ICA \\
		Spectral Methods & Spectral Clustering \\
		N-Body Methods & - \\
		Structured Grids & - \\
		Unstructured Grids & Belief Propogation \\
		MapReduce & Expectation Maximisation \\
		Combinational Logic & Hashing \\
		Graph Traversal & Bayesian Networks, DTs, \\
		 & Natural Language Processing \\
		Dynamic Programming & Forward-backward, inside-outside, \\
		 & variable elimination, value iteration \\
		Back-track and & Kernel Regression, \\
		Branch+Bound & Constraint Satisfaction, \\ 
		 & Satisficability \\
		Graphical Models & Hidden Markov Models \\
		Finite State Machine & - \\
		\bottomrule
	\end{tabular}
\label{table:MLDwarfs}
\end{table}

Not all the dwarfs are applicable to ML, and there is some overlap in applications, e.g., Dense and Sparse Linear Algebra both have utility for SVMs, PCA and ICA. These dwarfs pose an additional interesting consideration; conventional wisdom suggests that increased parallelisation can improve computation efficiency. However, as we have seen with the von Neumann bottleneck, one must also consider the movement of data itself, as well as the difficulty of a given computation. 

Fundamentally, most ML applications are simply collections of more general computational problems. The properties of novel devices and materials may thus have the potential to perform these computations in different and more efficient ways than existing algorithms. If we consider the fundamental ML computations in isolation, ignoring the minutiae of a given ML application, we may see that novel devices have wider potential applications than implementing a given ML approach. For example, if we consider the fact that ReRAM crossbars implement MVM, rather than a neural network, we see that they also have utility as linear algebra accelerators, giving them use in applications described by the first two of the Berkeley dwarfs. Jongerius et al.'s study of computational problems in popular benchmark suites~\cite{Jongerius2011} examines the proportion of the execution time taken up by a specific computational problem (e.g., sorting, matrix operations...). We suggest that targeted application of novel technologies as solutions to specific computational problems will have a significant impact in the context of ML accelerators. 

As with many engineering challenges, framing and contextualising the problem is only half the battle. Our analysis of memcomputing in Section~\ref{section:memcomputing} showed that what appears on the surface to be an elegant solution can end up falling down if one does not consider all the factors at play. Simply parallelising a given computation may not always result in a significant improvement in performance, as shown in work by Cabezas and Stanley-Marbell~\cite{Cabezas2011}, which presented a framework for understanding parallelism in terms of the distribution of instruction- and thread-level parallelism over the execution time and showed limitations in performance gains.

Many review articles in the devices literature discuss emerging technologies and their principles of operation, providing a comprehensive overview of the field, but many ultimately end with the message that useful implementations are far-distant due to engineering challenges and issues with scalability when compared to current CMOS devices. We instead suggest that emerging technologies may be better applied and commercialised if researchers shift their focus to the direct mapping of computational problems to the unique properties of new devices to achieve better performance in a given application than conventional digital CMOS devices using the above figures of merit.
\section{Summary and Conclusions}
\label{section:conclusions}
In the past decade, ML has become an important and widely used tool across a range of disciplines and fields. However, the fundamentally different nature of ML computations when compared to traditional computer programs has shown limitations with mainstream computing methods for these approaches. This has resulted in the development of a variety of proposals for novel materials, devices, and architectures. Despite the technical ingenuity, skilled manipulation of device and material properties and new physics that many of these experimental devices and approaches have demonstrated, the transfer from \textit{lab to fab} and the adoption of these devices by computer  architects is minimal at best. In this work, we have evaluated a variety of proposed new forms of computation and sought to explain this disconnect. Firstly, we provide an overview of the fundamental concepts of computational complexity and physical limitations of computing. We then provide a tutorial overview of ML, with a focus on representation. Next, we discuss the use of novel materials and devices as solutions to computational problems, alongside a brief overview of candidate technologies. Following this, we provide a survey of different approaches to ML accelerators, and compare the performance of different approaches. Our key findings are \textbf{(1)}: ANNs based on novel materials and devices significantly lag ANNs based on digital CMOS hardware (both fixed-function accelerator ASICs and programmable microprocessors, GPUs and FPGAs) in MNIST, a popular ML benchmark; \textbf{(2)}: NVM implementations of ANNs, despite having been explored for the best part of a decade, have demonstrated no real trend of progress and \textbf{(3)}: many proposed ML accelerators using experimental devices lack a physical realisation, relying on simulations of a system based on the performance of a few successfully fabricated devices. Though the problem of fabricating at scale is important to solve, suitable application of emerging technologies as solutions to computational problems, rather than as direct replacements for digital CMOS devices may circumvent this, as certain applications may not require high endurance or large numbers of devices. The first point in particular, the lag in performance, appears to be the result of differing conceptions of success between the materials, devices, and architectures communities. Finally, we propose several figures of merit and other suggestions that we hope will help unify the materials, devices, and architectures communities in the domain of ML research, enabling valid comparisons between approaches in both communities and fostering a dialogue to facilitate more impactful research.

\section*{Acknowledgements}
P. Stanley-Marbell is supported by an Alan Turing Institute award TU/B/000096 under EPSRC grant EP/N510129/1, by EPSRC grant EP/V047507/1, and by the UKRI Materials Made Smarter Research Centre (EPSRC grant EP/V061798/1). N.J. Tye acknowledges funding from EPSRC grant EP/L016087/1. S. Hofmann acknowledges funding from EPSRC (EP/P005152/1, EP/P007767/1). We also thank Jon Crowcroft for his comments and feedback on early drafts. Lastly, we acknowledge the contributions of James Meech and Jennifer Rodowicz in compiling the data in Figure 7.

% \pnasbreak splits and balances the columns before the references.
% If you see unexpected formatting errors, try commenting out this line
% as it can run into problems with floats and footnotes on the final page.
% \pnasbreak

\vspace{0.25in}

% Bibliography
\bibliographystyle{abbrv}
\bibliography{working-document}

\end{document}